\documentclass[final,5p,times,twocolumns,sort&compress]{elsarticle}

\usepackage{amssymb}
\usepackage{amsmath}

\newcommand{\TiV}{$^\text{nat}\text{Ti}(\text{p,x})^{48}\text{V }$}
\newcommand{\CuZn}{$^\text{nat}\text{Cu}(\text{p,x})^{65}\text{Zn }$}

\newcommand{\ZnIso}{$^{65}\text{Zn }$}

\newcommand{\MeV}[1]{\qty{#1}{\mev}}

\newcommand{\circlednum}[1]{\textcircled{\raisebox{-.9pt}{#1}}}
\usepackage{siunitx}
\usepackage{array} 
\usepackage{multirow} 
\usepackage{wrapfig}
\usepackage{tikz}
\usepackage{xcolor} 
\usepackage{caption}
\usepackage{subcaption}
\usepackage[hyphens]{url}
\usepackage[colorlinks=true,allcolors=blue]{hyperref}
\hypersetup{breaklinks=true}
\usepackage{colortbl}

\newcommand{\eq}{\, = \,}

\newcommand{\kev}{\mbox{ keV}}
\newcommand{\mev}{\mbox{ MeV}}

\usepackage{lineno}

\journal{Nuclear Instruments and Methods in Physics Research Section A}

\begin{document}

\begin{frontmatter}

\title{Beam Energy Measurement using a Bayesian Approach with the Stacked Foil Method}

\author[bern]{Alexander Gottstein\corref{cor1}}
\author[bern,insel,artorg]{Lorenzo Mercolli}
\author[bern]{Eva Kasanda}
\author[bern]{Isidre Mateu}
\author[bern]{Lars Eggimann}
\author[bern]{Elnaz Zyaee}
\author[bern]{Gaia Dellepiane}
\author[bern,napoli,infn]{Pierluigi Casolaro}
\author[bern,napoli]{Paola Scampoli}
\author[bern]{Saverio Braccini}

\cortext[cor1]{Corresponding author: alexander.gottstein@unibe.ch}



\affiliation[bern]{organization={Albert Einstein Center for Fundamental Physics (AEC), Laboratory for High Energy Physics (LHEP), University of Bern},
            addressline={Sidlerstrasse 5},
            city={Bern},
            postcode={3012},
            country={Switzerland}}

\affiliation[insel]{organization={Department of Nuclear Medicine, Inselspital, Bern University Hospital, University of Bern},
            addressline={Rosenbühlgasse 25},
            city={Bern},
            postcode={3010},
            country={Switzerland}}

\affiliation[artorg]{organization={ARTORG Center for Biomedical Engineering Research, University of Bern},
            addressline={Murtenstrasse 50},
            city={Bern},
            postcode={3008},
            country={Switzerland}}

\affiliation[napoli]{organization={Department of Physics ``Ettore Pancini'', University of Napoli Federico II},
            addressline={Complesso Univ. Monte S. Angelo},
            city={Napoli},
            postcode={80126},
            country={Italy}}

\affiliation[infn]{organization={INFN Sezione di Napoli},
            addressline={Complesso Univ. Monte S. Angelo},
            city={Napoli},
            postcode={80126},
            country={Italy}}

\begin{abstract}
We present a practical method to measure the energy of proton beams at a medical cyclotron using the stacked foil technique in combination with a Bayesian inference method. By measuring the $^{48}$V activity induced in a stack of irradiated titanium foils, the  proton energy can be inferred without relying on direct current or charge measurements, making the method suitable even for low-vacuum environments or air-exposed setups. This technique is further extended to configurations where the beam energy is degraded to levels around \MeV{8}. A Bayesian fit of the measured activity profile allows not only for a robust energy estimation but also for a consistent treatment of uncertainties and nuisance parameters. Monte Carlo simulations are employed to validate the underlying assumptions, including the impact of energy dispersion or cross-section uncertainties. Our results demonstrate that this method provides accurate beam energy measurements across several typical experimental setups used at the Bern Medical Cyclotron. Additionally, we evaluate the sensitivity of the method to the choice of nuclear cross-section data and assess how the number of foils in the stack affects the uncertainty in the inferred beam energy.
\end{abstract}


\end{frontmatter}

\section{Introduction} \label{s:intro}

Precise knowledge of proton beam energy is essential for a variety of research applications involving the irradiation of solid targets, such as in the production of novel radioisotopes for medical applications. This is becoming increasingly important as commercial production using solid targets is growing and requires accurate beam characterization for process optimisation and quality assurance. 

At the Bern Medical Cyclotron (BMC), where routine production of radiopharmaceuticals and academic research are conducted in parallel, precise beam energy determination plays a key role in a variety of different projects. Accurate knowledge of the actual delivered proton energy is crucial for cross-section measurements \cite{dellepiane_research_2021, dellepiane_optimised_2022, dellepiane_47sc_2022, dellepiane_cross-section_2022, braccini_optimization_2022}, where it enables more accurate predictions of reaction-product yields. Furthermore, it is also important for dosimetry in radiobiology studies (e.g.~in~\cite{constanzo_dosimetry_2019}) or radiation hardness tests (see \cite{anders_facility_2022, maia_oliveira_radiation-induced_2021, braccini_novel_2022}).
For proton beams in the \qty{20}{\mev} range, there are several methods that provide a reliable measurement of the beam energy, such as, for example, time-of-flight methods~\cite{ToFmethod}, the Rutherford backscattering method~\cite{CasolaroRutherfordBackScatter}, multi-leaf Faraday cup based methods~\cite{nesteruk_measurement_2019}, or methods making use of magnetic deflection of protons \cite{haffner_study_2019}, some of them explored by our group in the past. However, these methods require relatively complex experimental setups and are primarily suited for the measurement of the pristine beam energy in vacuum. 

As a simpler alternative, Burrage et al.~\cite{burrage_simple_2009} introduced the stacked-foil method, later refined by Asad et al.~\cite{asad_new_2015} or also Do Carmo et al. \cite{do_carmo_simple_2019}, which allows the beam energy to be inferred from induced radioactivity in a sequence of thin foils. Building on this methodology, we demonstrate that beam energy can be determined through activity measurements in the range of 8~–~\MeV{18} using combinations of titanium, copper, and niobium foils. In contrast to Asad’s approach, which relies on the $^\text{nat}\text{Cu}(\text{p,x})^{62,63,65}\text{Zn}$ reactions and activity ratios, we employ the well-established \TiV monitor reaction and analyse the activities directly, instead of considering activity ratios of the irradiated foils. A Bayesian inference framework is used to fit the measured activities against model predictions, naturally incorporating the non-linear dependence of the yields on the beam energy and providing a robust treatment of uncertainties. This enables a consistent quantification of the beam energy uncertainty through the marginalized posterior distribution. Furthermore, we show that the method remains applicable across different irradiation setups, and that the use of degrading elements and alternating foil materials extends its reach to energies below \MeV{10}.

In order to validate the main assumptions of our analysis, we compare the measurements and fit methodologies to full Monte Carlo (MC) simulations.

The presented method allows characterisation of the energy of a proton beam using a calibration-free irradiation setup paired with gamma spectroscopy using a High-Purity Germanium (HPGe) detector. The presented approach exploits the strong energy dependence of certain well-known reaction cross-sections induced by interactions between the beam and metallic foils. These reactions, referred to as monitor reactions, are commonly employed in the characterization of particle beams for radionuclide production, or to provide reference data in cross-section measurements~\cite{hermanne_reference_2018}. 

Monitor reactions are available for various energy thresholds, and together they span a broad energy range. This makes the stacked foil method well-suited for determining beam energies for protons that have already undergone significant energy degradation. 

For the measurements in this work in the range of 8~-~\MeV{19} and below, a combination of titanium, copper, and niobium foils was employed. Titanium and copper both produce well-characterized monitor reactions when irradiated by protons in the relevant energy range~\cite{hermanne_reference_2018}. The \TiV reaction allows for reliable monitoring mainly in the regions from 6~–~\MeV{9} and 14~–~\MeV{20}, where its cross-section is steepest. The \CuZn reaction, on the other hand, can provide the reference at lower energies in the range from 4~–~\MeV{6}. This is discussed in more depth in section~\ref{sec:calculations}.
Niobium foils were used as beam degraders, due to their relatively high stopping power. The beam energy was measured in multiple experimental configurations commonly used at the BMC for multidisciplinary research, including setups in air for which current measurements are not reliable.

\section{Materials and Methods}

The BMC, located at Inselspital, the Bern University Hospital, is used for the commercial production of radiopharmaceuticals as well as for multidisciplinary research activities. For the latter, the cyclotron (IBA 18/18 MeV Cyclone) is equipped with an external beam transfer line (BTL) on one of its eight exit ports, and an IBA Nirta solid target station (STS) on another. The other six outports are reserved for the commercial production of radiopharmaceuticals for clinical PET imaging.\newline The BTL is a \qty{6.5}{\meter}-long beam line that transports the proton beam from the cyclotron bunker to a separate experimental bunker that can be accessed independently. It is equipped with two sets of quadrupole magnets for beam focusing and a pair of XY-steering magnets for beam alignment. This setup allows flexible beam delivery for different experiments. The beam profile can be focused down to about one millimetre or defocused to produce a quasi-flat beam. A wide range of beam currents can be achieved, from approximately \qty{150}{\micro\ampere} down to \qty{1.5}{\pico\ampere} \cite{auger_low_2015}, supporting experimental research spanning from detector development to cross-section measurements.

To determine the energy of the beam delivered by the cyclotron, measurements were performed in different experimental setups using the stacked foil method. This technique relies on irradiating a stack of metal foils, which become activated through the production of radionuclides, and the yield $Y$ depends on the beam energy and the cross-section. The primary reactions of interest for this method are the productions of $^{48}\text{V}$ and $^{65}\text{Zn}$ through the well-known monitor reactions $^\text{nat}\text{Ti}(p,x)^{48}\text{V}$ and \CuZn.
The expected product yield $Y$ in each foil depends on the entry beam energy $E_\text{in}$, the foil thickness $d_i$, the mass stopping power $S_p$ of the foil material, and the cross-section $\sigma$ of the monitor reaction. Given these parameters, the expected activity $A_{exp}$ produced from irradiating a foil with an integrated current $Q$ can be calculated analytically as a function of the initial beam energy $E_\text{in}$ using the following equation:
\begin{align} \label{eq:A_layer}
A_\text{exp} &= Q \cdot Y(E_\text{in}) \\ &= Q \cdot \frac{ (1 - \exp(-\lambda \cdot t_i))}{t_i} \cdot \frac{N_A \cdot \eta}{m_{mol} \cdot q} \int_{E_\text{out}}^{E_\text{in}} \frac{\sigma(E)}{S_p(E)} dE \qquad \left[ Bq \right]
\label{eq:A_layer_2}
\end{align}
Here, $E_\text{in}$ and $E_\text{out}$ are the energies at which the proton beam enters and exits the respective layer of the foil stack, $N_A$ is the Avogadro constant, $\eta$ is the stoichiometric number, $m_{mol}$ is the molar mass of the target material, $\lambda$ is the decay constant of the reaction product of interest, $q$ is the charge of the projectile (+\qty{1}{\elementarycharge} for protons), and $Q$ is the integrated current, which is the total charge with which the layer of titanium is irradiated, i.e. $Q = \int_0 ^{t_i} I(t) \,dt$. For irradiation times $t_i$ that are much shorter than the half-life of the product of interest, the saturation factor $\left( 1 - \exp({-\lambda \cdot t_i}) \right)\cdot t_i^{-1}$ reduces to the decay constant $\lambda$. 

We assume that the loss in fluence due to the beam divergence and the nuclear reactions as the beam travels through the stack of foils is small and can be neglected. This is motivated by the fact that the size of the foils is much bigger than the size of the collimated beam spot. This assumption is already underlying Eq.~\eqref{eq:A_layer}, as we approximate the convolution of the cross-section and the differential particle fluence as the product of the cross-section and the total charge. Furthermore, this enables the computation of the energy $E_\text{out}$ for the $i^\text{th}$ foil in the stack by 
\begin{equation}
    E_\text{out} \approx E_\text{in} - S_p(E_\text{in}) \cdot d_i \;, 
\end{equation}
where the stopping power, $S_p(E_\text{in})$, is extracted from the simulation package SRIM~\cite{ziegler_srim_2010}. An accurate measurement of $Q$ is not always possible, such as in low-vacuum environments in which ionisation of air can influence the measured current. To reliably characterize the proton beam energy in these conditions, we treat the integrated current $Q$ as a free parameter and therefore do not need to rely on a current or charge measurement to determine the beam energy.

\begin{figure*}[h]
  \centering
  \includegraphics[width=\textwidth]{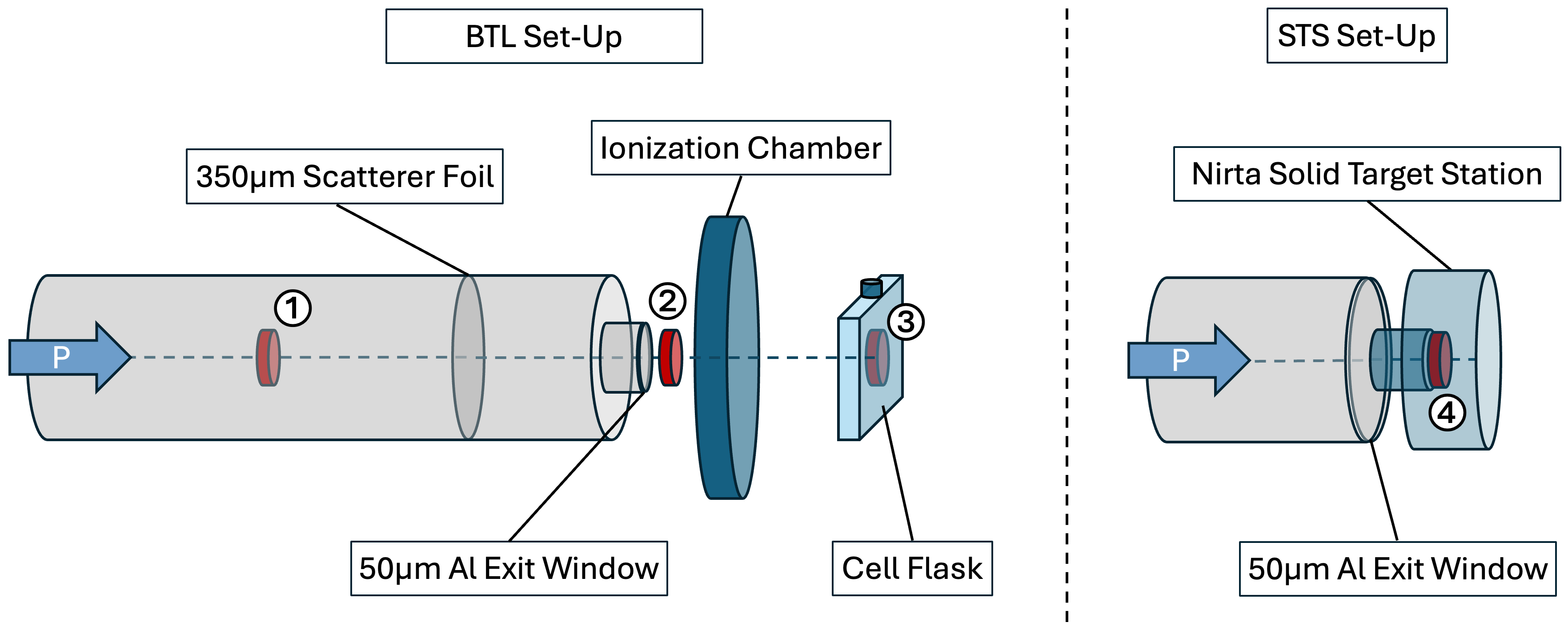}
  \caption{A sketch of the different positions where the beam energy has been measured for the BTL set-up. The coin holding the foil stack has been placed at the positions indicated in red. At the first indicated position, the pristine beam energy is measured (\raisebox{.5pt}{\textcircled{\raisebox{-.9pt} {\scriptsize 1}}} \textit{BTL Energy}). At the second position, the beam energy has been degraded by an aluminium scatterer and an aluminium extraction window (\raisebox{.5pt}{\textcircled{\raisebox{-.9pt} {\scriptsize 2}}}~\textit{After Scatterer}). At the third position, the beam energy is measured after having additionally passed through air, an ionisation chamber and a cell flask's wall (\raisebox{.5pt}{\textcircled{\raisebox{-.9pt} {\scriptsize 3}}} \textit{Cell level}). \raisebox{.5pt}{\textcircled{\raisebox{-.9pt} {\scriptsize 1}}} and \raisebox{.5pt}{\textcircled{\raisebox{-.9pt} {\scriptsize 2}}} have been performed using the target holder shown in Figure~\ref{fig:target holder}. In another measurement (\raisebox{.5pt}{\textcircled{\raisebox{-.9pt} {4}}} \textit{STS Energy}) the beam energy of another beamline outport of the BMC was characterized. On this beamline a commercial Nirta solid target station (STS) is installed in combination with a \qty{50}{\um} Ti exit window.}
  \label{fig:btl_scheme}
\end{figure*}

\subsection{Beam Line Setups}

The beam energy was measured for both available outports at the BMC, the BTL outport, and the outport where the STS is installed. Despite the protons being accelerated by the same machine for both the BTL and the STS outports, the pristine beam energy extracted from the two ports can differ slightly. This discrepancy can be attributed to factors such as local magnetic field inhomogeneities and beam extraction parameters at the respective outport, e.g. the radial position of the stripper and its angle (see Figure 7 of Ref. \cite{haffner_study_2019}). Furthermore, the extracted beam delivered to each outport exhibits slightly different characteristics, such as variations in emittance and transverse phase-space distribution. To evaluate the robustness of our methodology under these differing beam conditions, measurements were performed at both outports. Beyond methodological considerations, we were also interested in determining the beam energy specific to each extraction line. Accordingly, measurements were performed at both outports.

On the BTL, energy measurements were performed in three different configurations, which are illustrated in Figure~\ref{fig:btl_scheme}. Configuration \raisebox{.5pt}{\textcircled{\raisebox{-.9pt} {\small 1}}} represents the energy of the pristine beam extracted from the BTL port of the cyclotron. Configuration \raisebox{.5pt}{\textcircled{\raisebox{-.9pt} {\small 2}}} represents the energy of the beam after it has been extracted into air through a \qty{350}{\um}-thick aluminium beam scatterer and a \qty{50}{\um}-thick aluminium beam extraction window. This configuration is most often used for low-flux irradiations. Finally, configuration \raisebox{.5pt}{\textcircled{\raisebox{-.9pt} {\small 3}}} represents the energy of the beam that is delivered to cells in the context of in-vitro radiobiology experiments performed using the BTL (in preparation). All the measurements on the BTL have been performed with a stripper angle of 91\textdegree \ of \mbox{stripper~\#\! 1} on carrousel~4, i.e. the rotating holder inside the cyclotron on which the stripper foils dedicated to the BTL outport are mounted  \cite{haffner_study_2019}.

At the STS outport, only the measurement in configuration \raisebox{.5pt}{\textcircled{\raisebox{-.9pt} {4}}} took place. There, the pristine beam energy was measured at the level of the STS, where the pristine beam has already passed through a \qty{50}{\um} titanium beam extraction window and the beam energy was decreased slightly. The measurement with the STS has been performed with a stripper angle of 99.8\textdegree \ of the stripper \#\! 1 on carrousel~7.

\subsection{Target Holder and Target Station}

\begin{figure}[h]
\centering
\includegraphics[width=1.0\linewidth]{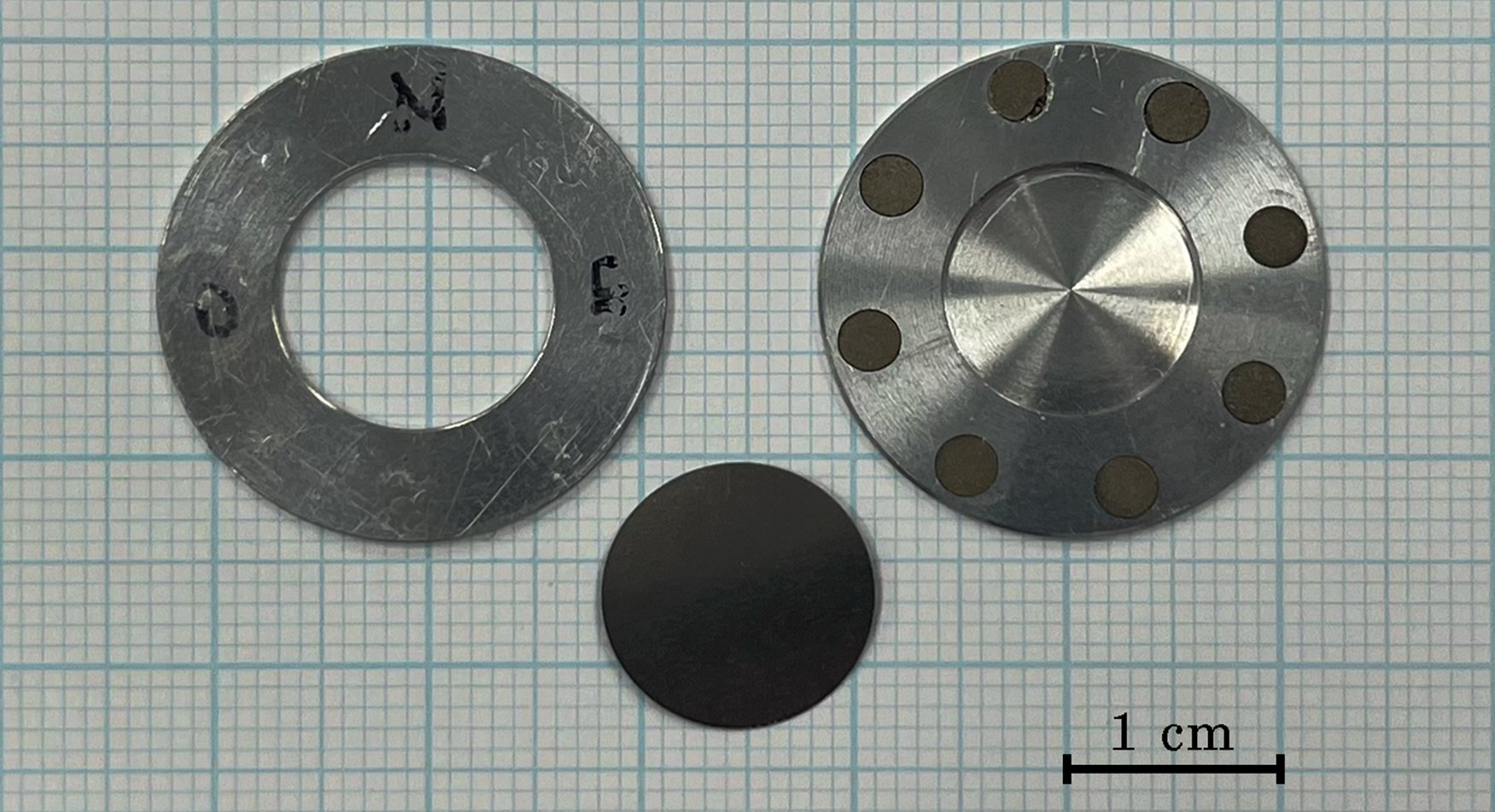}
\caption{The target coin has a lid (left) with a circular opening, attached to its backside (right) with small magnets. Between the two a single titanium foil is shown.}
\label{fig:target coin}
\vspace{-6mm}
\end{figure}

\begin{figure*}[h]
    \centering
    \includegraphics[width=0.65\linewidth]{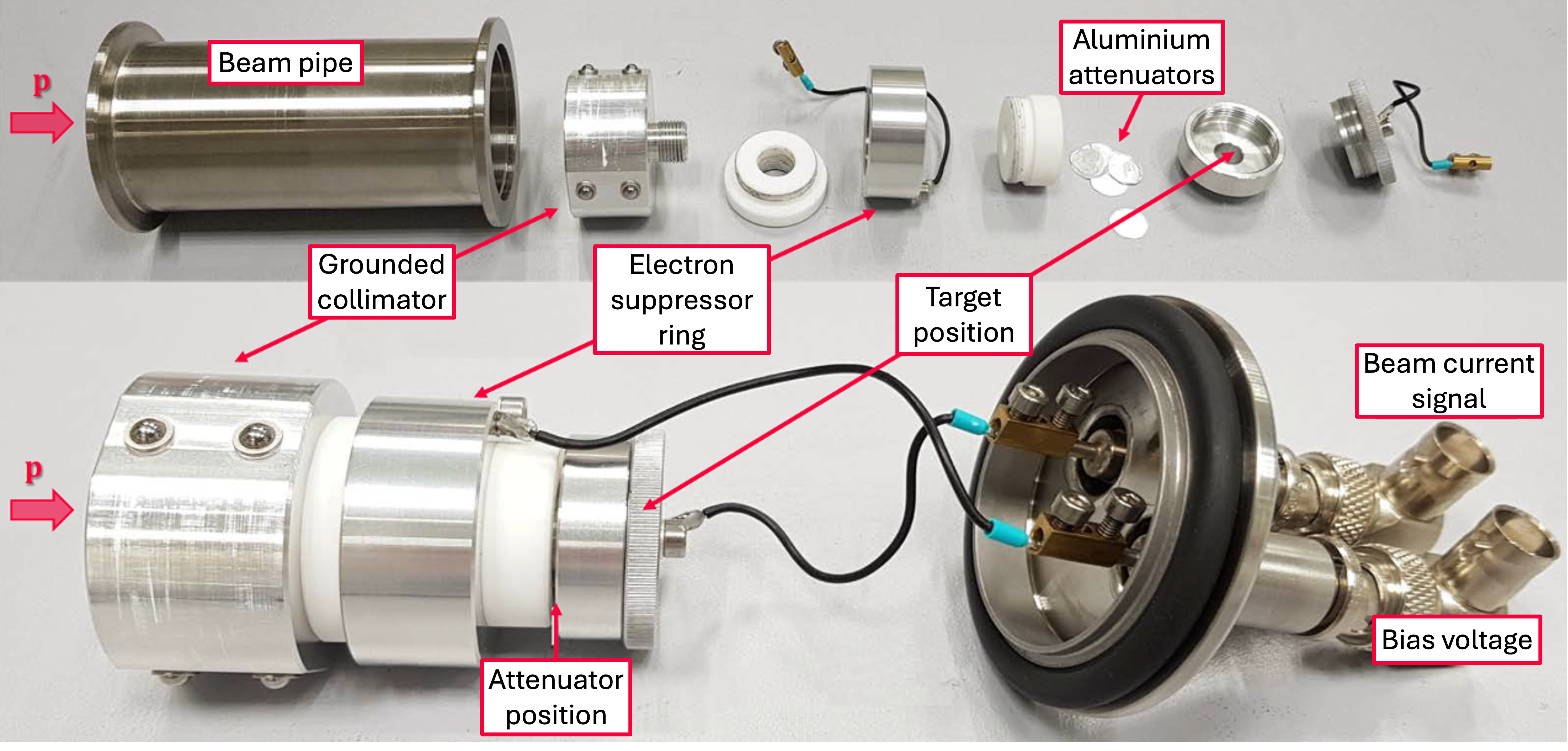}
    \caption{The target station that was used for measurements \raisebox{.5pt}{\textcircled{\raisebox{-.9pt} {\scriptsize 1}}} and \raisebox{.5pt}{\textcircled{\raisebox{-.9pt} {\scriptsize 2}}}. The incident proton beam (indicated with labelled arrow) hits the target foil stack at the target position after being collimated by the grounded collimator and passing by the electron suppressor ring, which is only needed if one is interested in a target on current measurement. The depicted aluminium attenuators are optional, depending on the desired experimental configuration. The current measurement and bias voltage components have not been used for the measurements in this publication. The picture is adapted with modifications from \cite{dellepiane_research_2021}.}
    \label{fig:target holder}
\end{figure*}

For all energy measurements the target holder shown in Figure~\ref{fig:target coin} was used. The coin-shaped target holder served to secure the stacked foils in place during the irradiations. The used coin featured an aperture marginally smaller than \qty{13}{\mm} in its lid, providing sufficient mechanical support for the stacked foils while ensuring that the incoming proton beam remained undegraded. \\
For the energy measurements on the BTL the target station depicted in Figure~\ref{fig:target holder} was utilized to keep the target holder in place during the irradiation. The integrated collimator of the target station limited the beam spot on the foil stack to an \qty{8}{\mm} diameter, preventing proton loss from beam scattering or partial obstruction.
For the beam energy measurement in configuration \raisebox{.5pt}{\textcircled{\raisebox{-.9pt} {\small 3}}} the target station was installed after the ionization chamber, in its own vacuum independent from the beam line's main vacuum. Gafchromic films were used to verify the beam alignment. 
For the energy measurement on the STS the target coin holding the foil stack was irradiated using a commercial NIRTA solid target station~\cite{dellepiane_new_2022}.

\subsection{Foil Stacks}\label{sec:foil_stacks}

The main component of the irradiated target is the foil stack. Stacked foils of Ti, Cu, and Nb with a diameter of \qty{13}{\mm} are embedded within the target holder. The Ti and Cu foils have a thickness of \qty{25}{\um}, whereas the Nb foils are \qty{125}{\um} thick. 
To minimize  thickness variations, foils of the same material were laser-cut from a single parent sheet.

\begin{figure}[h]
    \centering
    \includegraphics[width=\linewidth]{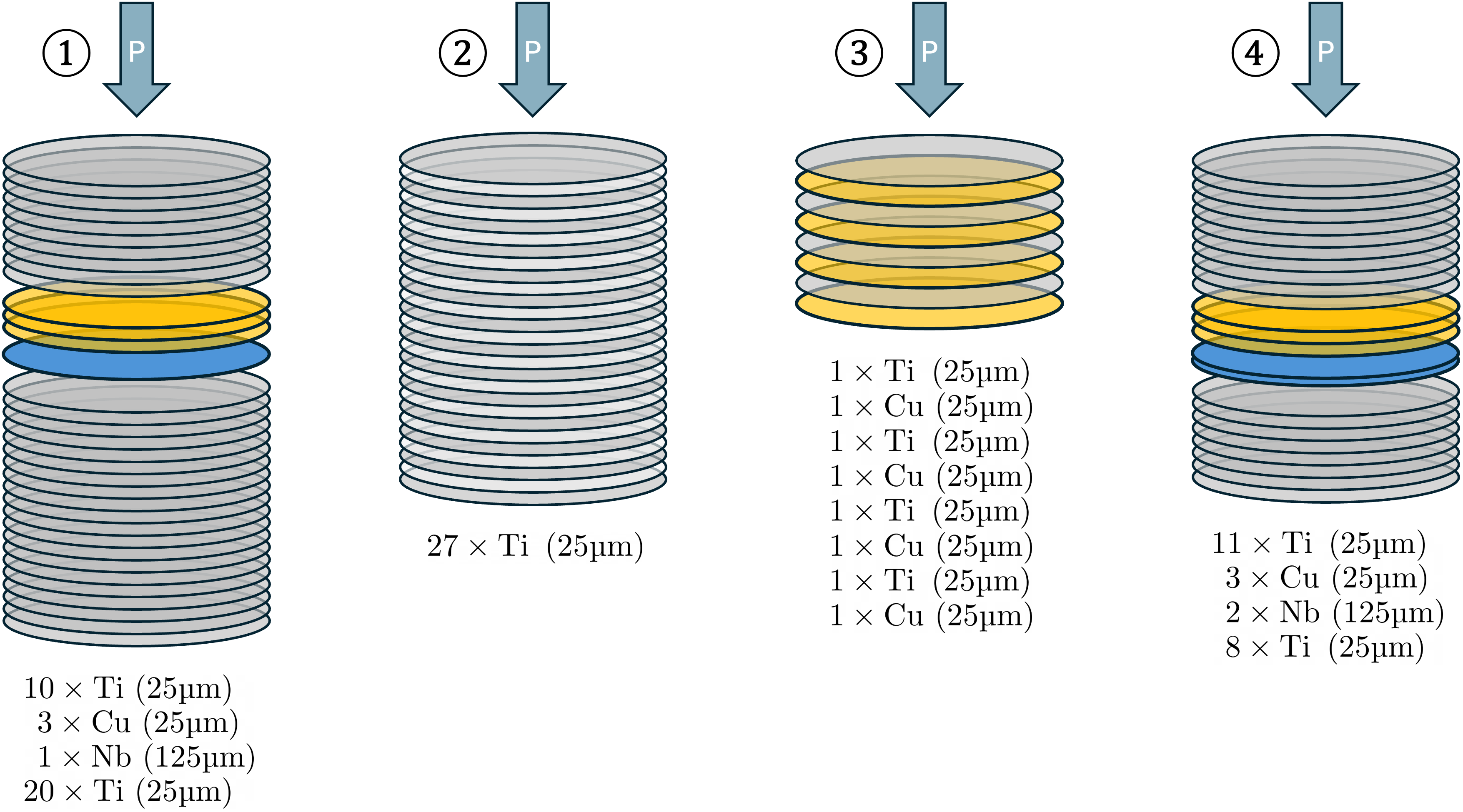}
    \caption{The composition of all four foil stacks. \raisebox{.5pt}{\textcircled{\raisebox{-.9pt} {\scriptsize 1}}} shows the composition of the 'pristine BTL energy' measurement's foil stack, \raisebox{.5pt}{\textcircled{\raisebox{-.9pt} {\scriptsize 2}}} the 'After Scatterer' measurement's composition, \raisebox{.5pt}{\textcircled{\raisebox{-.9pt} {\scriptsize 3}}} the 'Cell Level' measurement's composition, and \raisebox{.5pt}{\textcircled{\raisebox{-.9pt} {4}}} the 'STS measurement' composition of the foil stack. The direction of the incoming proton beam is indicated by the arrows. The different colours of the layers indicate the material of the layer, where grey is indicating titanium, yellow is for copper and blue is for niobium. In stack \raisebox{.5pt}{\textcircled{\raisebox{-.9pt} {\scriptsize 2}}} the two shades of grey indicate active layers (dark) and beam degrading layers (pale).}
    \label{fig:stack details}
\end{figure}

The Nb foils are used as beam energy degraders to shift the proton energy towards the more sensitive region of the excitation function, the Cu foils can be used as both beam energy degraders and active measurement foils. Alongside the cross-section of the \TiV reaction, the cross-section of the \CuZn reaction can be used as a secondary indicator. The specific arrangements of stacked foils for the performed measurements are shown in Figure~\ref{fig:stack details}.

\begin{figure}[htb]
  \centering
  \includegraphics[width=\linewidth]{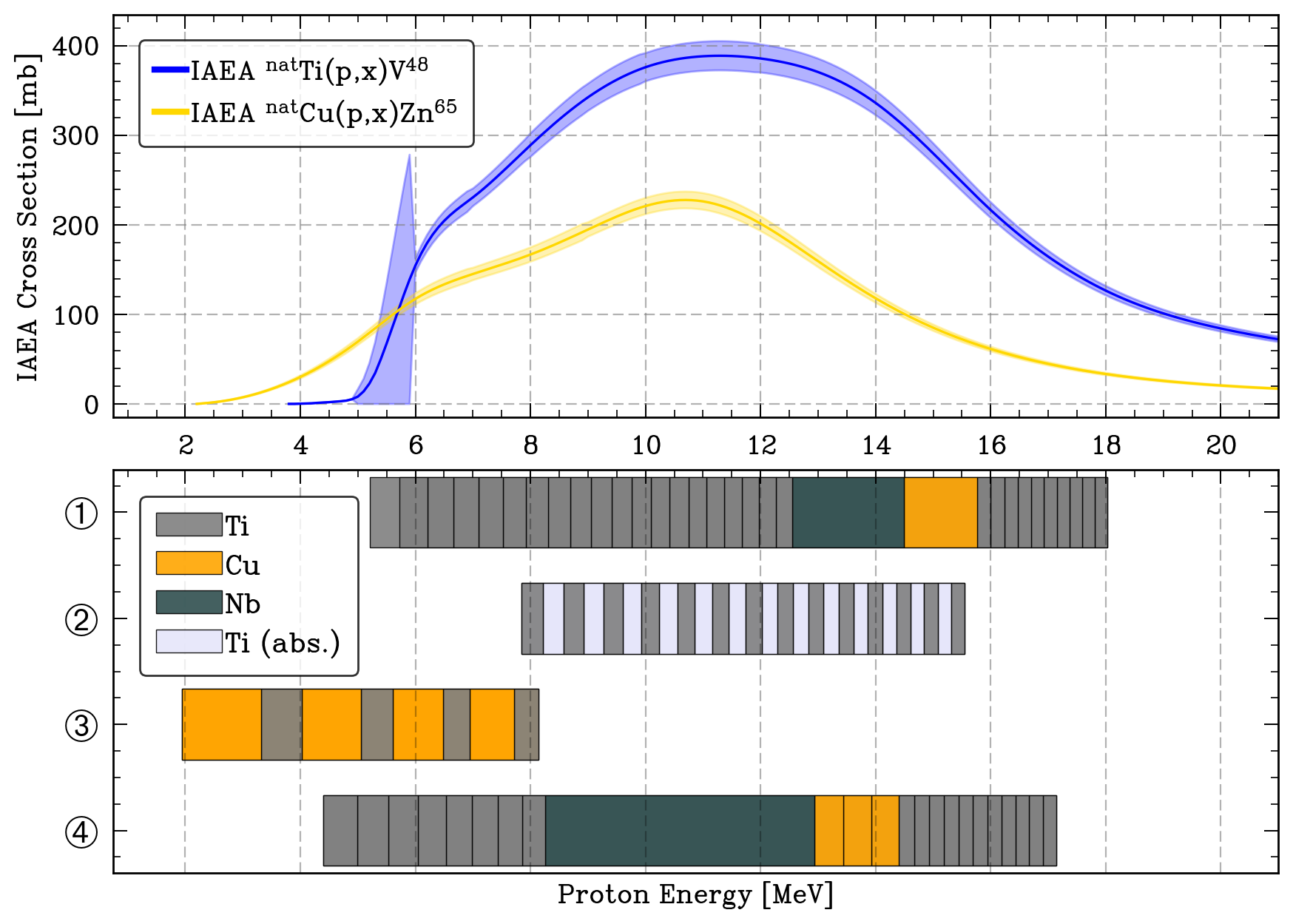}
  \caption{The above plot shows the IAEA-recommended cross-sections (\citet{hermanne_reference_2018}) for the \TiV reaction in blue and the \CuZn reaction in yellow. The bottom plot shows the expected foil energies for each measurement’s foil stack. These energies are calculated from the best fit initial energies listed in Table \ref{tab:energy results}.}  
  \label{fig:crosssections_and_stacks}

\end{figure}

\subsection{Activity Measurement} \label{sec:activity measurement}
The activities of the irradiated foils were determined using a commercial gamma spectrometer that consists of a \mbox{Mirion/Canberra} N-type GR2019 wide energy band high purity germanium detector (HPGe) \cite{noauthor_mirion_2025} and a \textit{LYNX-MCA Digital Signal Analyzer}. The spectrometer's efficiency was determined using a multi-gamma source beforehand. The activity of the foils was measured using the HPGe spectrometer. By taking into account the elapsed time between the end of beam (EOB) on target and the HPGe measurement and knowing the respective decay constants, the activity of the foils at the EOB was then determined. To evaluate the $^{48}$V activity we looked at the \qty{1312}{\kev} $\gamma$-emission (98.2~\% branching ratio) from the $^{48}$V decay, as well as the \qty{1116}{\kev} $\gamma$-emission for the \ZnIso decay (50.04~\% branching ratio).

As will be shown in Section~\ref{s:results}, the measured activity data strongly constrains the fit of the energy of the proton beam~$E_0$. Therefore, we consider also situations with less data points, i.e.\ with a reduced number of foils, and study the impact on the uncertainty of $E_0$.

\subsection{Determining the Beam Energy from the Measured Activities} \label{sec:calculations}

We rely on a Bayesian fitting procedure to determine $E_0$ at the entry of the first foil. There are several advantages over a standard frequentist fitting methodology, in particular given the highly non-linear model of Eq.~\eqref{eq:A_layer}. Standard frequentist fitting algorithms struggle to reliably estimate the covariance matrix for models that involve integrals and strongly correlated parameters, as in Eq.~\eqref{eq:A_layer_2}. This leads to an underestimation of the parameter uncertainties, which we want to avoid. Furthermore, we are primarily interested in the value of $E_0$ for the experimental set-ups described in Section~\ref{sec:foil_stacks}. All the other parameters, such as the integrated current $Q$, the foil thickness $d$, $E_\text{in/out}$ in different foils, are not relevant for our purposes. The Bayesian framework allows us to treat these nuisance parameters consistently and to account for their uncertainty in the marginalized posterior distribution of $E_0$. Finally, the Bayesian sampling of the posterior distributions gives us control over correlation among the fitting parameters. 

A drawback of the Bayesian approach is the need for considerable computational resources. The posterior distribution needs to be sampled and from Eq.~\eqref{eq:A_layer} it is clear that every single sample point requires a numerical integration. Fitting $E_0$ for a single data set can be computationally demanding. 

\begin{figure}[h]
    \centering
    \includegraphics[width=1.0\linewidth]{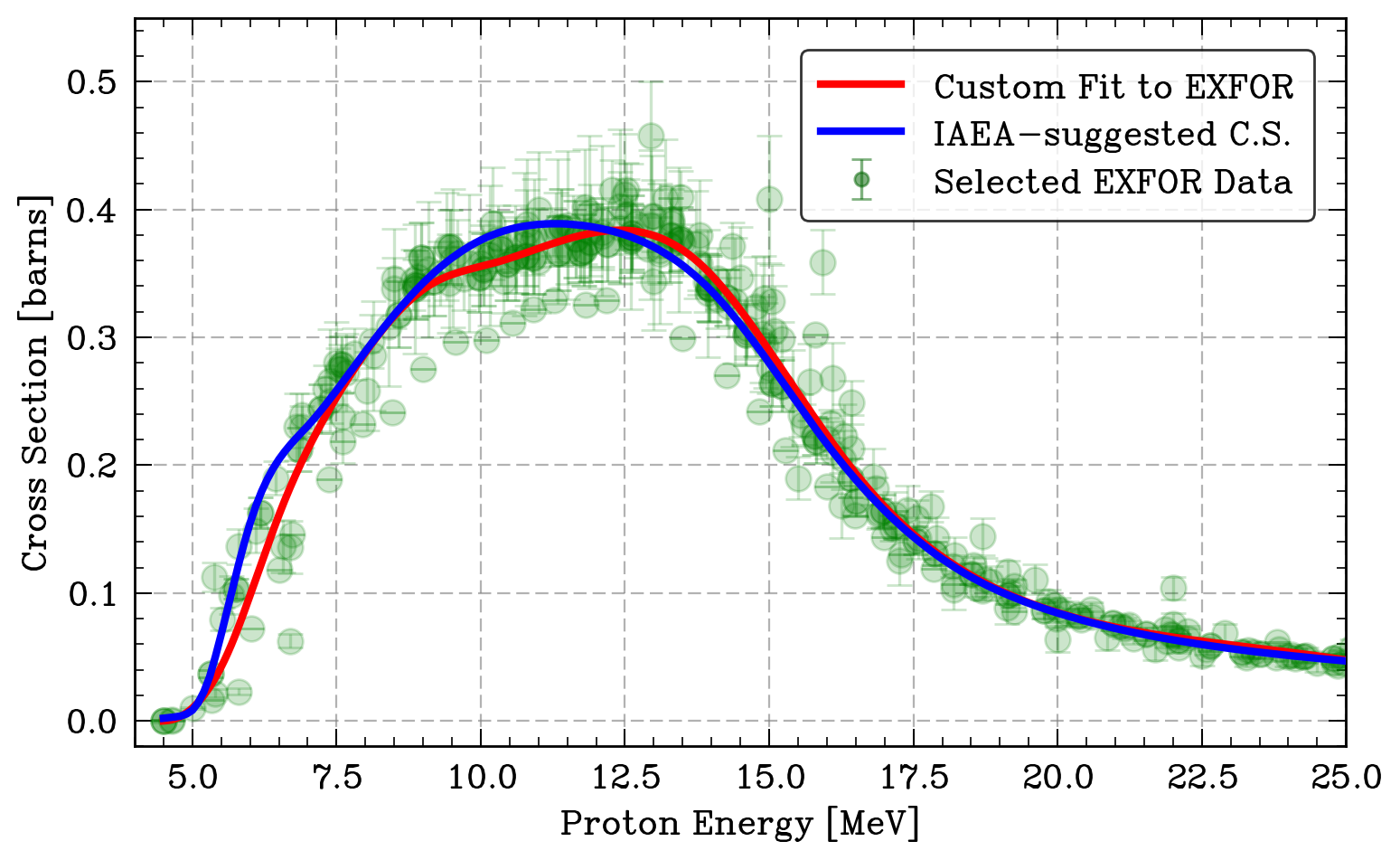}
    \caption{Comparison of IAEA-suggested cross-section for the \TiV reaction (blue line) and a custom spline fit (red line) to selected EXFOR data (green data points) for the same reaction cross-section. The custom fit to the EXFOR data is similar to what is suggested in~\cite{cervenak_new_2020}.}
    \label{fig:cross section comparison}
\end{figure}

To set up the Bayesian model, we assume a Gaussian likelihood. 
Although the measured activities follow a Poisson distribution, the count rates are sufficiently high to justify a Gaussian approximation. For the initial beam energy $E_0$, we assume a flat but not completely uninformative prior distribution. Depending on the beam line elements that are installed before the stack of foils, we can already restrict the possible values of the energy. Also, given our cyclotron's geometry and parameters of control, we know that the beam energy cannot exceed \MeV{19}. Therefore, we choose relatively uninformed the prior distributions of $E_0$ as 

\begin{equation} \label{eq:prior_E}
E_0 \sim 
\begin{cases}
    \text{Uniform}(12~\mathrm{MeV},\, 19~\mathrm{MeV}) & \text{for  \circlednum{\small 1}, \circlednum{\small 2}, and \circlednum{\small 4},} \\[1.2ex]
    \text{Uniform}(6~\mathrm{MeV},\, 19~\mathrm{MeV}) & \text{for \circlednum{\small 3}.}
\end{cases}
\end{equation}
As shown in Section~\ref{s:results}, the fit results have virtually no dependence on the priors in Eq.~\eqref{eq:prior_E}.
For the sake of numerical stability, the time-integrated current $Q$ is split into a fitting parameter $P$ and a fixed numerical factor of $1 \cdot 10^{-5} \, \mbox{C}$, which sets the scale of the total charge on target. Also for $Q$, we choose a uniformly distributed distribution
\begin{equation}
    Q \eq 1 \cdot 10^{-5} \, \mbox{C} \, \cdot \, \mathcal{X} \quad \mbox{with} \quad \mathcal{X} \, \sim \, \mbox{Uniform}(0,10) \,. 
\end{equation}
The stopping power and the reaction cross-section are crucial inputs for the calculation of the yield in a single foil. We use the International Atomic Energy Agency (IAEA) monitor cross-section from~\citep{hermanne_reference_2018} for the \TiV and \CuZn reactions, as shown in Figure~\ref{fig:cross section comparison}. The relative error on these cross-sections quoted by IAEA is around 5~\%. However, the literature on the \TiV reaction suggests that the error might be larger, e.g. around 10\%~\cite{cervenak_new_2020}. In particular, around the dipole resonance, the shape of the cross-section might be different. Figure~\ref{fig:cross section comparison} presents the reference cross-section along with our fit to the cross-section data extracted from EXFOR (see \cite{otuka_towards_2014}). In order to account for the uncertainty in the cross-section, we apply a scaling factor $\Delta \sigma_{Ti, \, Cu}$ to the IAEA cross-sections for \TiV and \CuZn that introduces a normally distributed variation of 10~\%, i.e. the prior distributions for the two scaling factors are independently sampled for the two target materials
\begin{equation}
    \Delta \sigma_\text{Ti, Cu}  \, \sim \, \mathcal{N}(1, 0.1) \; . 
\end{equation}

To verify the method, it was made sure that the results for the posterior distribution of $E_0$ agree within the statistical uncertainty when we use our own fit for the \TiV cross-section rather than the IAEA cross-section (see results in Table~\ref{tab:cs comparison}). This will become apparent in the results in section~\ref{s:results}.

In the energy range and for the materials relevant to this study, the SRIM stopping powers \cite{ziegler_srim_2010} are well known. We added an overall scaling factor $\Delta S_p$ for the stopping powers of Cu, Ti and Nb with a 5~\% Gaussian prior
\begin{equation}
    \Delta S_{p}  \, \sim \, \mathcal{N}(1, 0.05) \;.  
\end{equation}

It should be noted that this factor is sampled independently for each material. A bin-wise error parameter on stopping power and cross-section would not be meaningful, as they are a smooth function of the proton energy. 
From Eq.~\eqref{eq:A_layer} it is clear that also the foil thickness $d$ may have a significant impact on the uncertainty of $E_0$. The prior on $d$ is a Gaussian distribution with a 15~\% standard deviation, i.e.
\begin{equation}
    d \, = \, \left\{ \begin{array}{l} \qty{25}{\um} \; \mbox{for Ti and Cu} \\ \qty{125}{\um} \; \mbox{for Nb} \end{array} \right\} \cdot \Delta d \quad \mbox{with} \quad \Delta d \; \sim \; \mathcal{N}(1, 0.15) \,. 
\end{equation}
As the foils are laser-cut from a common parent sheet, the prior on $d$ is sampled only once and applied to all foils in the stack.

We observe a good convergence of the fit. The measurement data are strongly constraining the posterior distribution on $E_0$ and reduces the prior dependence of the fit result to a minimum. The sampling of the posterior distributions relies on a No-U-Turn algorithm \cite{hoffman2014} and its implementation in the software packages Julia~\cite{Julia}, Turing.jl~\cite{fjelde_turingjl_2025} and ArviZ~\cite{kumar_arviz_2019}. 
Since we use a Gaussian likelihood, we cross-checked our results with a frequentist $\chi^2$ fit to the data. As expected, the mean values are the same within the statistical error, while the statistical uncertainty on $E_0$ is slightly underestimated in the frequentist fit. 

\begin{figure}[h]
    \centering
    \includegraphics[width=\linewidth]{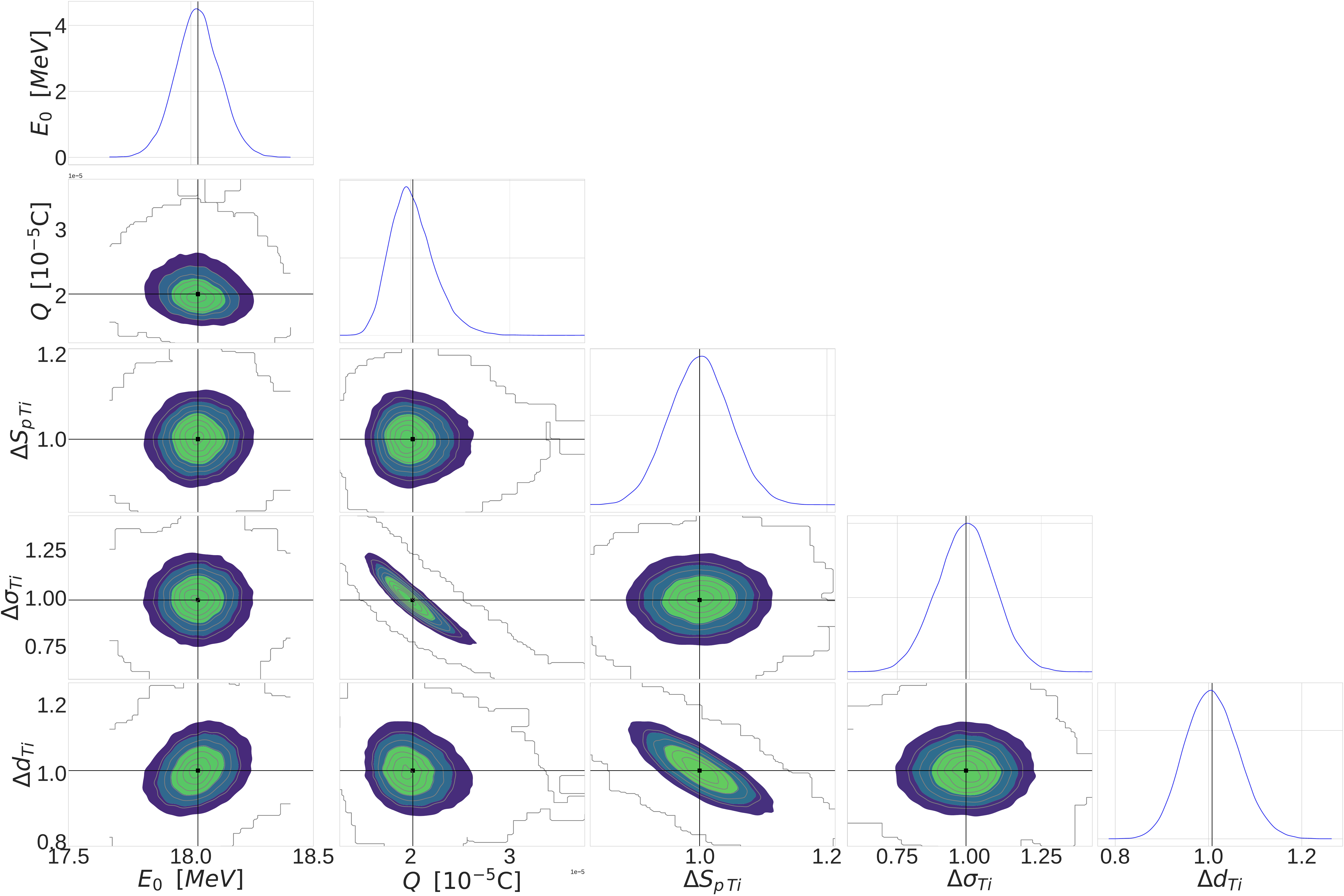}
    \caption{Pair plot for the maximum energy measurement in the BTL. For better visualization, the plot shows only $\Delta S_p$ and $\Delta \sigma$ of titanium. }
    \label{fig:pairplot}
\end{figure}

Figure~\ref{fig:pairplot} shows an exemplary pair plot for measurement \raisebox{.5pt}{\textcircled{\raisebox{-.9pt} {\small 1}}}. Overall, there is little correlation among the posterior distributions of $E_0$ with the other variables. The strong negative correlations between $\Delta\sigma_\text{Ti}$ and $Q$ as well as between $\Delta d_\text{Ti} $ with $\Delta S_{p \, \text{Ti}}$ follow directly from~Eq.~\eqref{eq:A_layer_2}.

\begin{figure}[htb]
    \centering
    \includegraphics[width=1.0\linewidth]{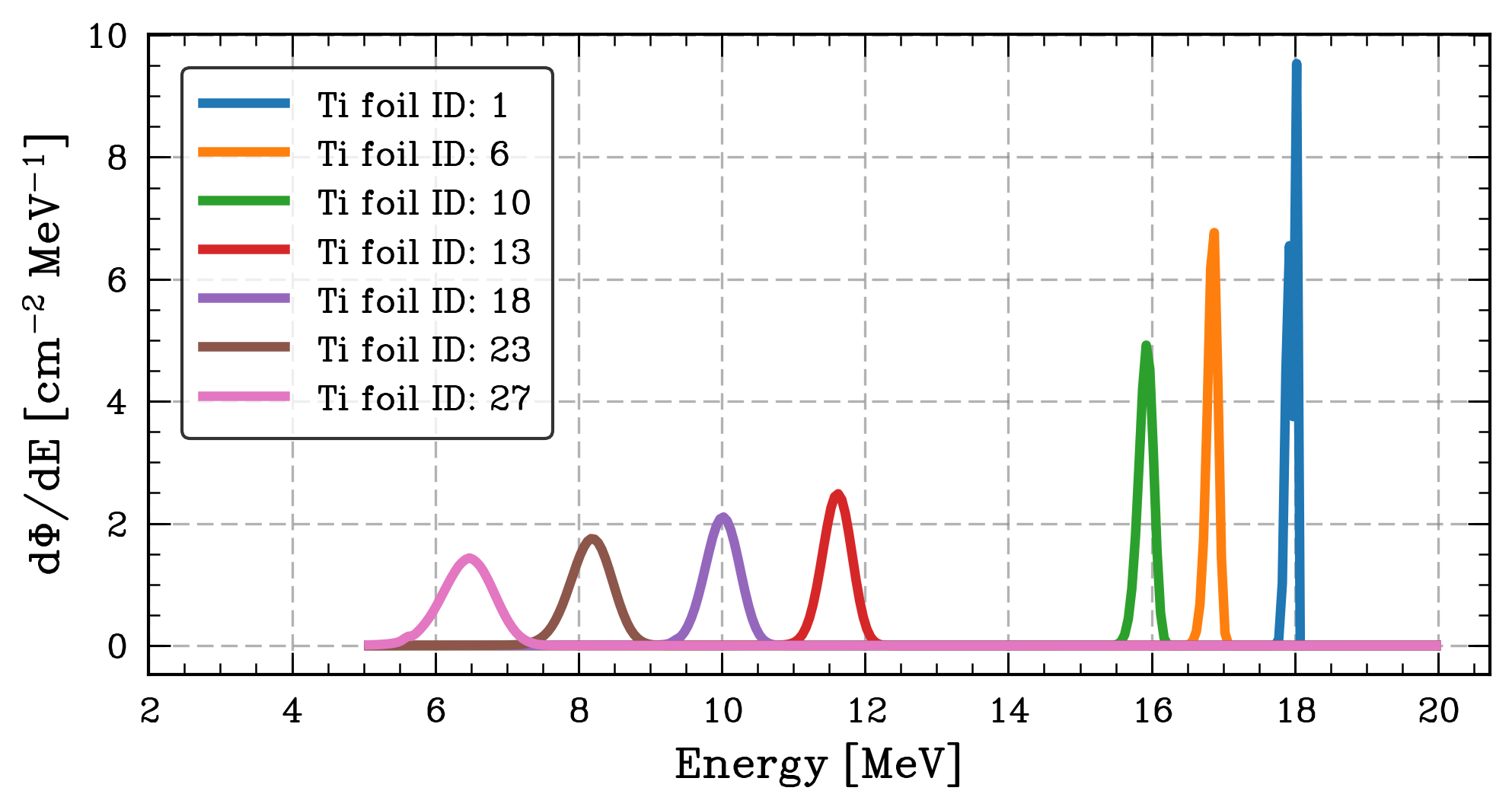}
    \caption{The simulated differential proton fluence in selected foils for the target of measurement \raisebox{.5pt}{\textcircled{\raisebox{-.9pt} {\scriptsize 1}}} illustrates the energy dispersion as the beam travels through the stack of Ti foils. The initial energy is $E_0 = $~\MeV{18.03} and the energy bins are \qty{50}{\kev}. For better visualisation we only show a selected amount of foils.}
    \label{fig:fluka_energy_dispersion_fluence}
\end{figure}

The analysis presented in this section assumes mono-energetic protons in each layer of the stack. Naturally, the pristine beam is not strictly mono-energetic, and energy dispersion occurs as it passes through the successive target layers. In order to justify the simplified mono-energetic model of Eq.~\eqref{eq:A_layer} we performed a set of Monte Carlo particle transport simulations with FLUKA ~\cite{battistoni_overview_2015,ahdida_new_2022,donadon_flair3_2024}. Figure~\ref{fig:fluka_energy_dispersion_fluence} illustrates that even with a perfectly mono-energetic initial beam, there is a significant broadening of the differential fluence. 

\begin{figure}[htb]
    \centering
    \includegraphics[width=1.0\linewidth]{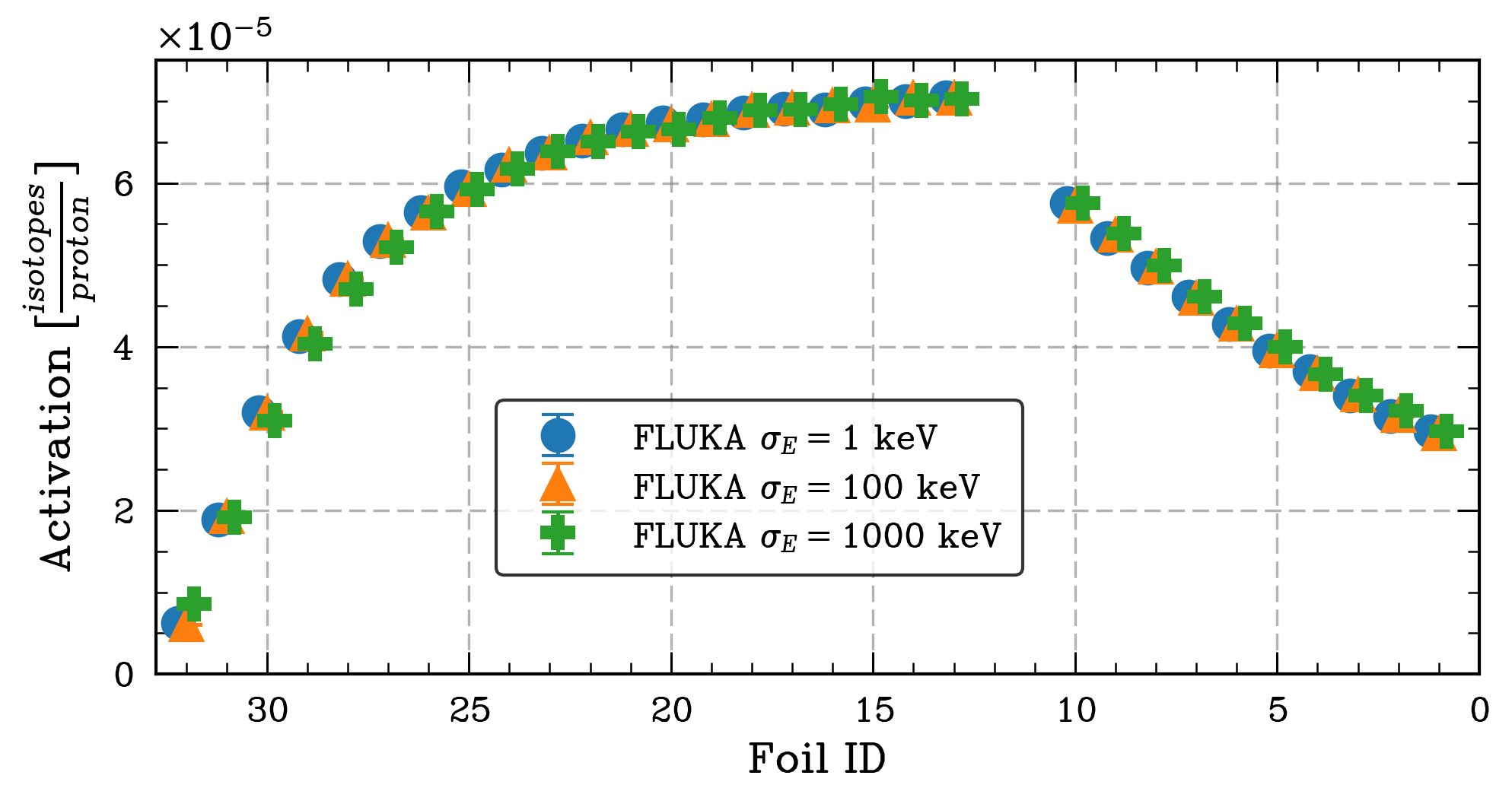}
    \caption{Activation products in the Ti layers for the target in \raisebox{.5pt}{\textcircled{\raisebox{-.9pt} {\scriptsize 1}}} from FLUKA simulations with different widths of a Gaussian energy profile, i.e. $\sigma_E = $~\qty{1}{\kev}, \qty{100}{\kev}, and \qty{1}{\mev}. The initial energy is $E_0 = $~\MeV{18.03} and the energy bins are \qty{50}{\kev}.}
    \label{fig:fluka_energy_dispersion}
\end{figure}

To further point out the validity of the method, we can show that the beam dispersion has only a marginal impact on the yield in each layer. Figure~\ref{fig:fluka_energy_dispersion} shows the $^{48}\text{V}$-activation of the Ti-foils for the set-up in \raisebox{.5pt}{\textcircled{\raisebox{-.9pt} {\small 1}}}. The initial beam was modelled with a Gaussian energy profile with different widths $\sigma_E = $~\qty{1}{\kev}, \qty{100}{\kev}, and \qty{1}{\mev}.\footnote{

For the cyclotron under consideration, the energy gain per orbit is $\Delta E = $~\qty{64}{\kev}. This value is obtained following the calculation of~\citet{braccini_particle_2013}, assuming an average magnetic field of $B = $~\qty{1.4}{\tesla} and an RF voltage of  $V_{RF} = $~\qty{32}{\kilo \volt}. As the stripping efficiency is nearly \qty{100}{\percent} for H$^{2-}$ ions at \MeV{18}, we can assume that protons from only two orbits are extracted at the same time from a single stripper foil, leading to an extracted energy spread on the order of $\Delta E$ as calculated above.}
The activation changes marginally, despite an energy dispersion that spans three orders of magnitude. This difference is much smaller than the measurement error of the activity in the Ti layers. This is likely due to the fact that, except for energies near the cross-section threshold, there are no very steep gradients in the cross-section and the stopping power. The integral over the energy in Eq.~\eqref{eq:A_layer_2} washes out any sensitivity to the energy spread of the beam. It is therefore justified to rely on the mono-energetic beam model of Eq.~\eqref{eq:A_layer}. It is important, though, to keep in mind that the statistical error which is quoted for $E_0$ does not correspond to any spread in energy of the beam. Rather, it is the statistical uncertainty of the central beam energy. 

From the FLUKA simulations, we can also conclude that with the experimental setup and target design under consideration, we will never be sensitive to the energy dispersion of the proton beam. 

\section{Results and Discussion}\label{s:results}

\subsection{Energy Measurement Results}

\noindent Using the methodology outlined above, the optimal energy values yielding the best agreement with experimental measurements were determined, as presented in Table~\ref{tab:energy results}. The posterior distribution for the beam energy is mostly Gauss-shaped, i.e.\ the highest density intervals (HDI) agree with the standard deviation quoted as uncertainty. For the two cases `After Scatterer'~\raisebox{.5pt}{\textcircled{\raisebox{-.9pt} {\small 2}}} and `Cell Level'~\raisebox{.5pt}{\textcircled{\raisebox{-.9pt} {\small 3}}} the results are compared to simulated results, where the energy degradation throughout the known respective components has been computed using the software package \texttt{LISE++}~\cite{tarasov_lise_2008}).

 \begin{table*}[htbp]
     \centering
     \vspace{4mm}
     \begin{tabular}{|c||c|c|c|c|c|}
          \hline
          \rowcolor[gray]{0.9} \textbf{Measurement} & \textbf{Energy $E_0$} & \textbf{Std Dev} & \textbf{HDI}$_{2.5\%}$ & \textbf{HDI}$_{97.5\%}$  & \textbf{Simulation} \\
          \hline
          BTL Energy \raisebox{.5pt}{\textcircled{\raisebox{-.9pt} {\small 1}}} & \qty{18.03}{\mev} & ±\,\qty{0.09}{\mev} & \qty{17.84}{\mev} & \qty{18.20}{\mev} & – \\
          \hline
          After Scatterer \raisebox{.5pt}{\textcircled{\raisebox{-.9pt} {\small 2}}} & \qty{15.54}{\mev} & ±\,\qty{0.12}{\mev} & \qty{15.29}{\mev} & \qty{15.78}{\mev} & 15.65 MeV\\
          \hline
          Cell Level \raisebox{.5pt}{\textcircled{\raisebox{-.9pt} {\small 3}}} & $\,\,$\qty{8.14}{\mev} & ±\,\qty{0.29}{\mev} & $\,\,$\qty{7.58}{\mev} & $\,\,$\qty{8.71}{\mev} & $\,\,$\MeV{7.81}\\
          \hline
          STS Energy \raisebox{.5pt}{\textcircled{\raisebox{-.9pt} {4}}} & \qty{17.14}{\mev} & ±\,\qty{0.13}{\mev} & \qty{16.89}{\mev} & \qty{17.40}{\mev} & – \\
          \hline
     \end{tabular}
     \vspace{3mm}
     \caption{Best fit parameters from the Bayesian model with associated uncertainties. For the two latter BTL scenarios \raisebox{.5pt}{\textcircled{\raisebox{-.9pt} {\scriptsize 2}}} and \raisebox{.5pt}{\textcircled{\raisebox{-.9pt} {\scriptsize 3}}} simulation results are included, which take the 'BTL Energy' from \raisebox{.5pt}{\textcircled{\raisebox{-.9pt} {\scriptsize 1}}} as a starting point and calculate the energy degradation using the software package \texttt{LISE++} \cite{tarasov_lise_2008}.}
     \label{tab:energy results}
 \end{table*}
\begin{figure*}[htbp]
    \centering
    \begin{subfigure}{0.45\textwidth}
        {\includegraphics[width=\linewidth]{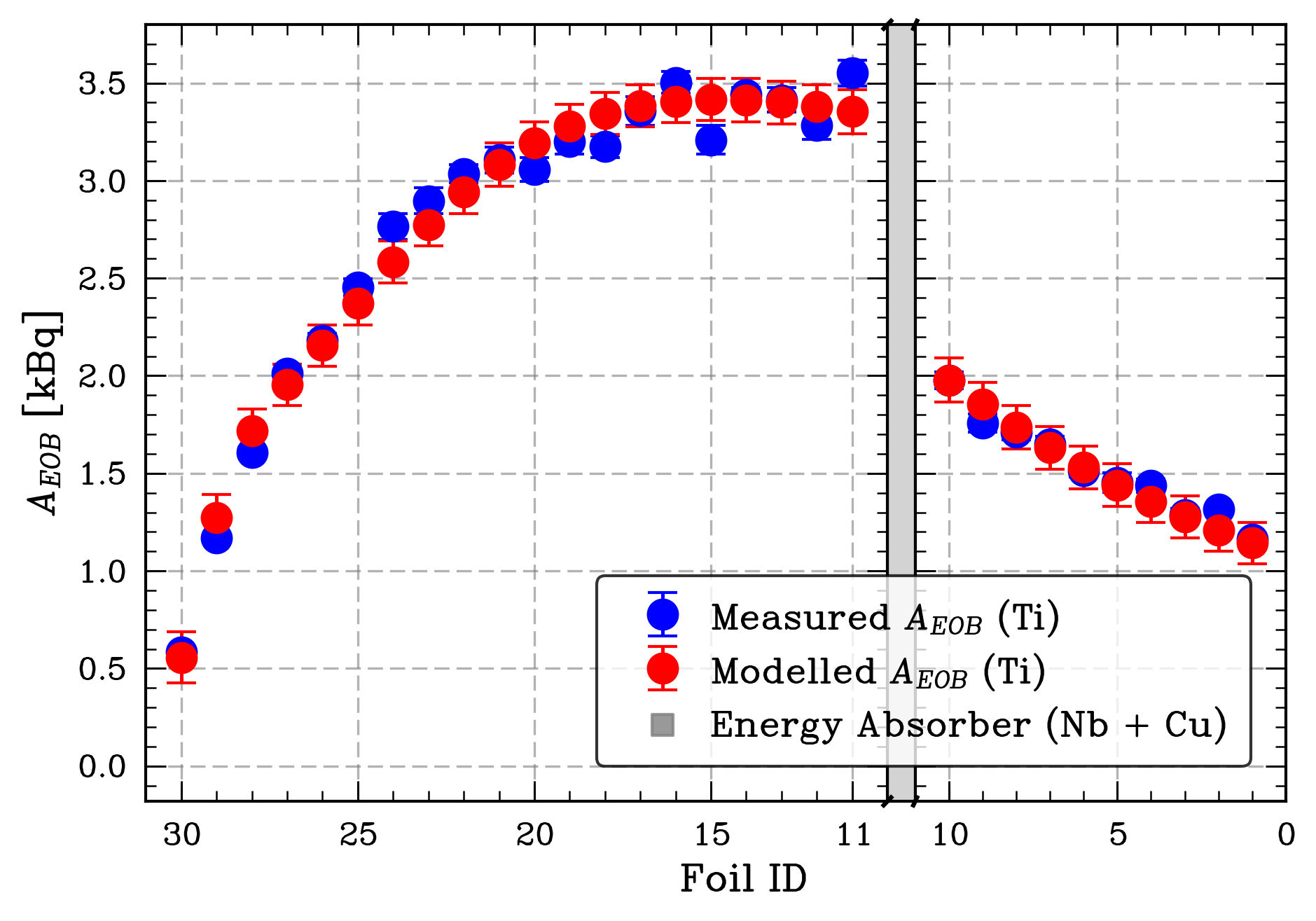}}
        \caption{'Pristine BTL energy' \raisebox{.5pt}{\textcircled{\raisebox{-.9pt} {\scriptsize 1}}}: Measured activities versus model predictions using the best fit energy of $E_0~=~$~\MeV{18.03}. The grey bar schematically indicates the energy absorbing Cu/Nb layers.}
        \label{fig:sub1}
    \end{subfigure}
    \hspace{1cm}
    \begin{subfigure}{0.443\textwidth}
        {\includegraphics[width=\linewidth]{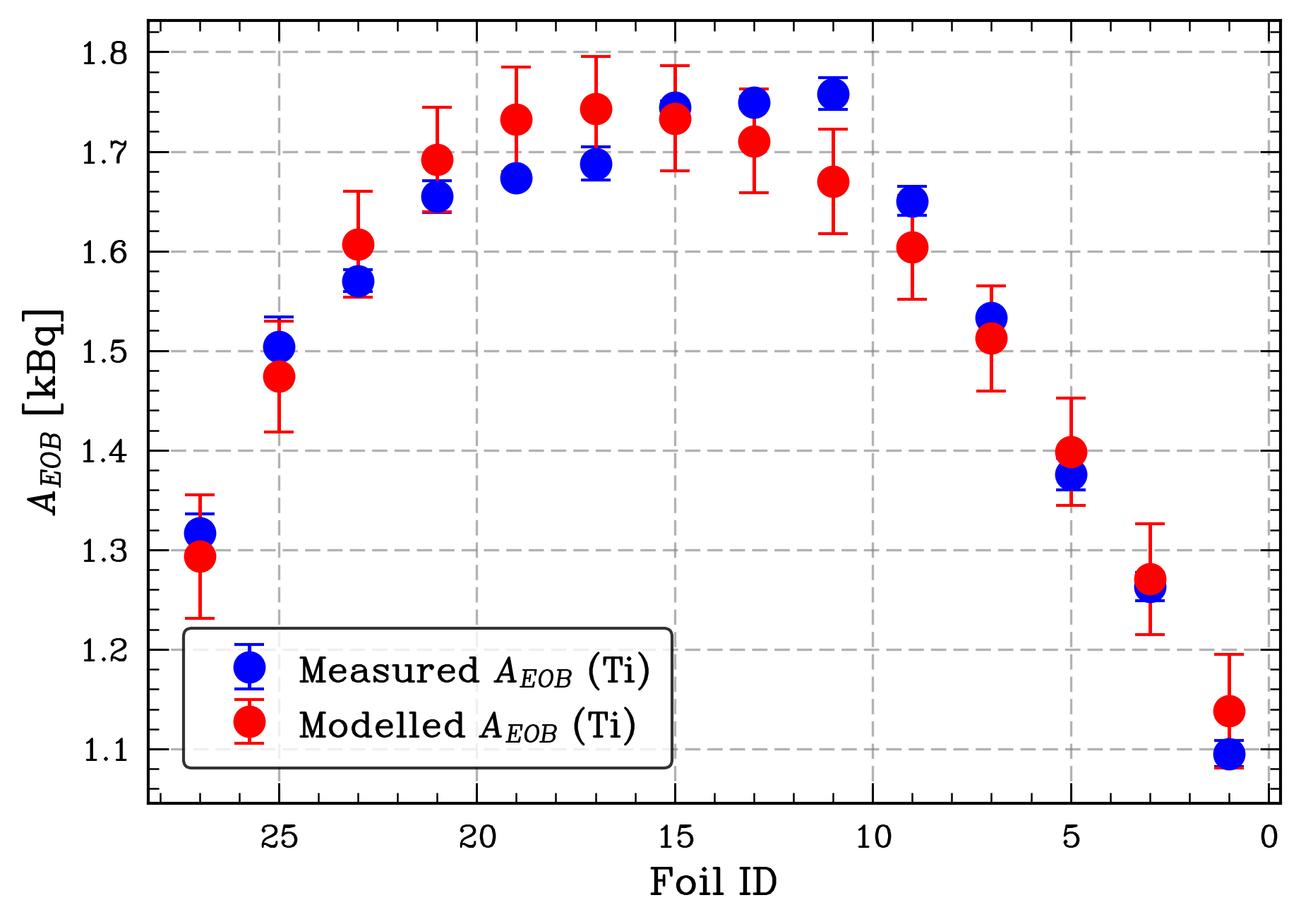}}
        \caption{'After Scatterer' measurement \raisebox{.5pt}{\textcircled{\raisebox{-.9pt} {\scriptsize 2}}}: Measured activities versus model predictions using the best fit energy of $E_0~=~$~\MeV{15.54}. In this case only odd foils have been used for the measurement.}
        \label{fig:sub2}
    \end{subfigure}
    
    \vspace{0.1cm} 

    \begin{subfigure}{0.45\textwidth}
        {\includegraphics[width=\linewidth]{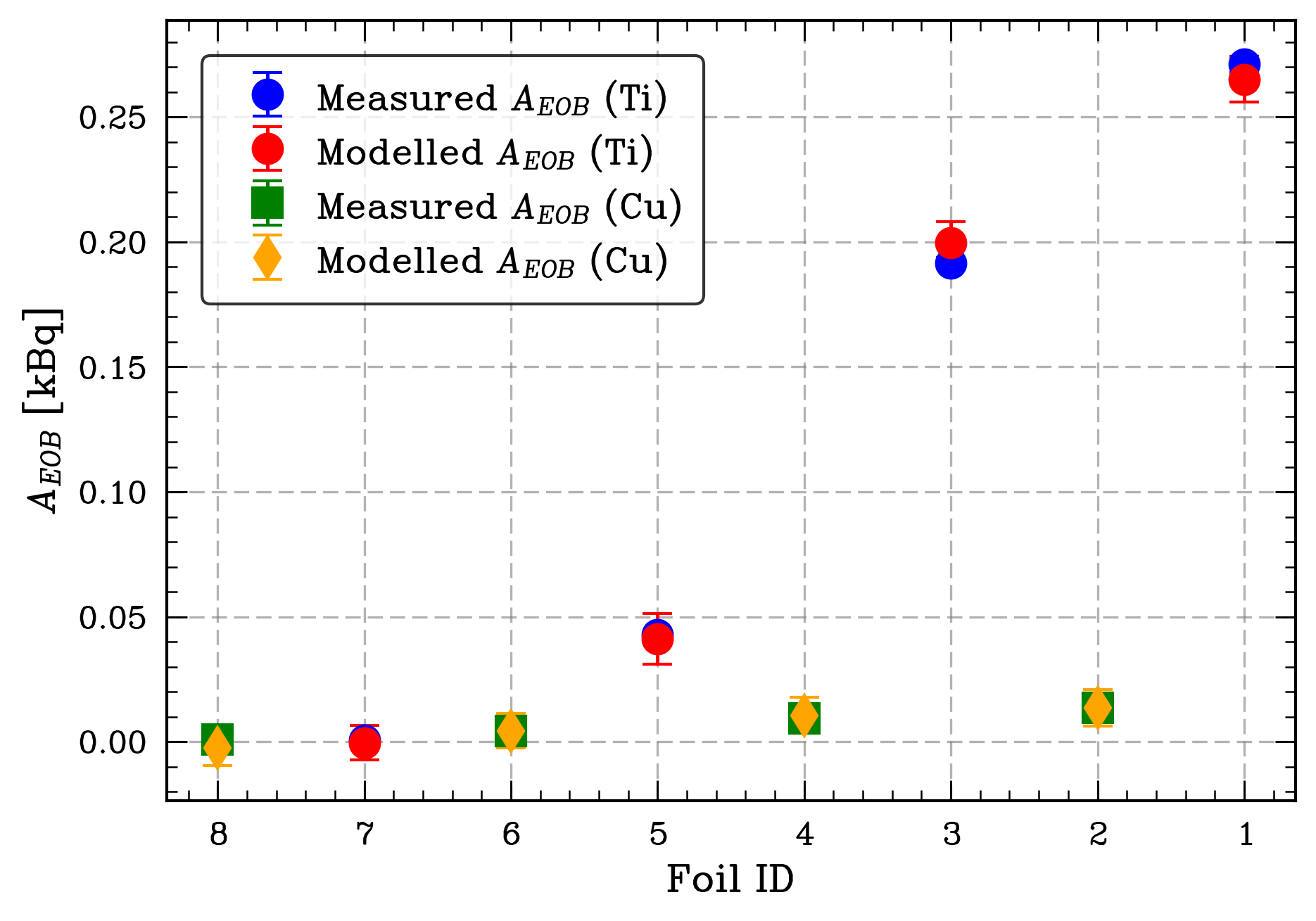}}
        \caption{`Cell Level' measurement \raisebox{.5pt}{\textcircled{\raisebox{-.9pt} {\tiny 3}}}: Measured Ti and Cu activities versus the respective model predictions using the best fit energy of $E_0~=~$~\MeV{8.14}.\\}
        \label{fig:sub3}
    \end{subfigure}
    \hspace{1cm}
    \begin{subfigure}{0.45\textwidth}
        {\includegraphics[width=\linewidth]{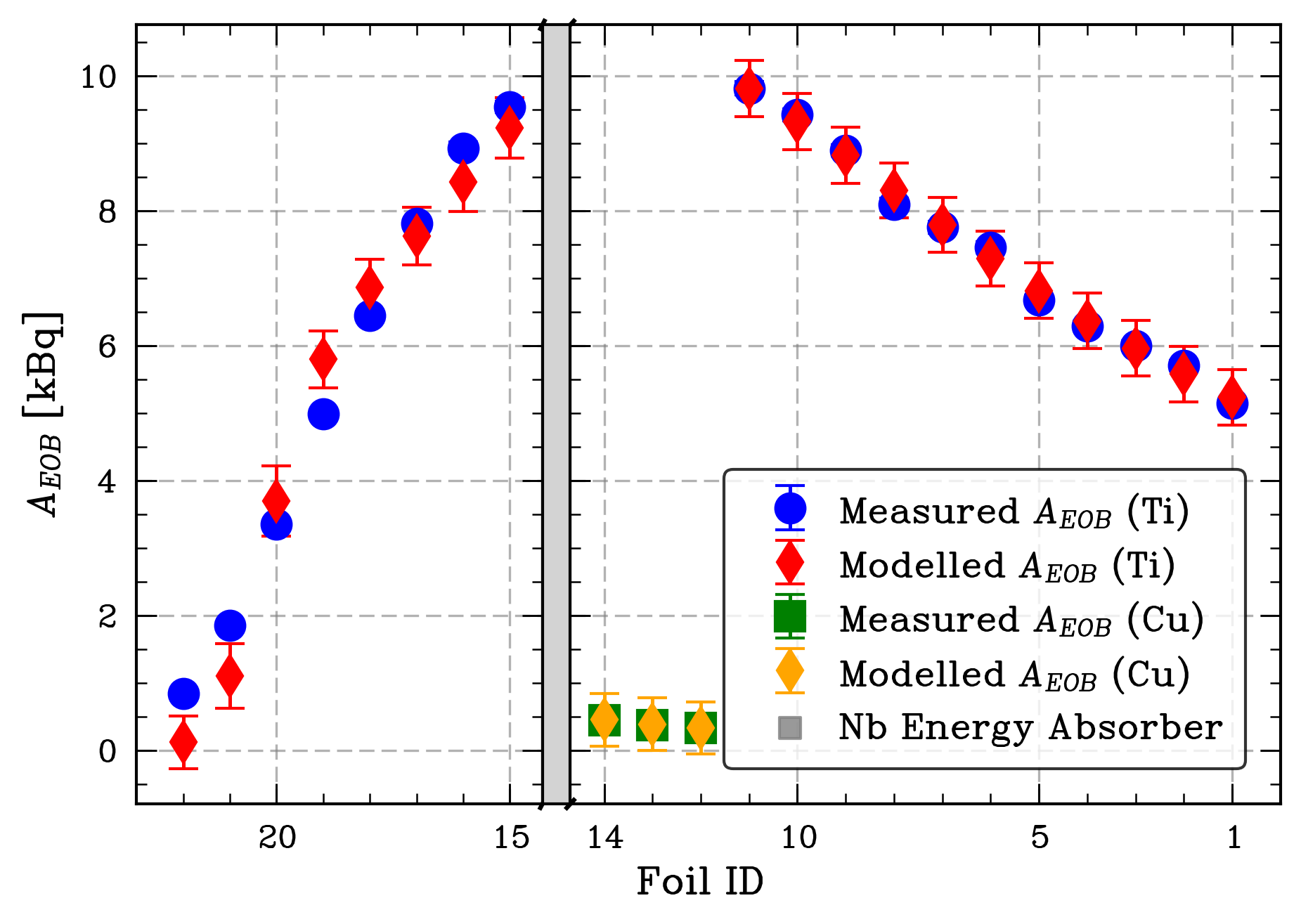}}
        \caption{'Pristine STS' energy \raisebox{.5pt}{\textcircled{\raisebox{-.9pt} {4}}}:
        Measured Ti and Cu activities versus model predictions using the best fit energy of $E_0~=~$~\MeV{17.14}. The grey bar schematically indicates the energy absorbing Nb layers.}
        \label{fig:sub4}
    \end{subfigure}

    \rule{\linewidth}{0.4pt}
    \vspace{0.01cm}
\begin{subfigure}{0.47\textwidth}
        \includegraphics[width=\linewidth]{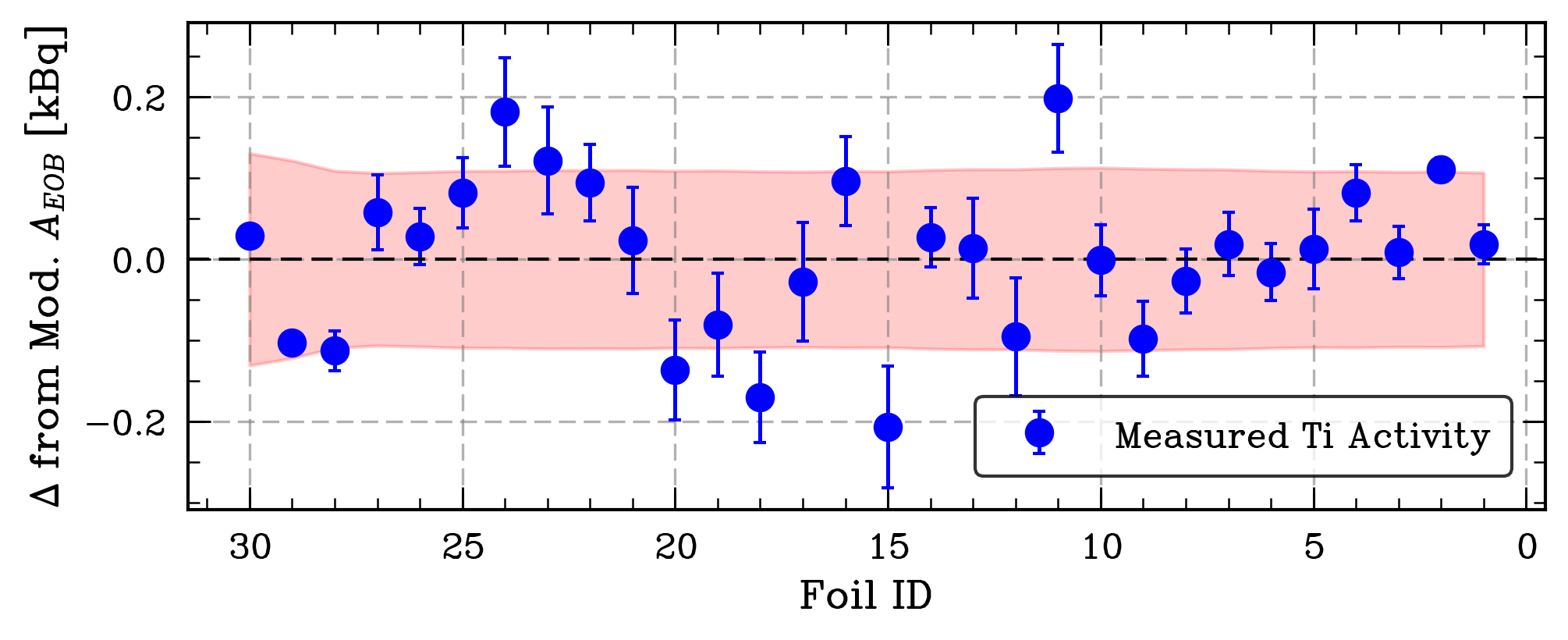}
        \caption{Residual plot for (a).} 
        \label{fig:sub1_res}
    \end{subfigure}
    \hspace{0.5cm}
    \begin{subfigure}{0.47\textwidth}
        \includegraphics[width=\linewidth]{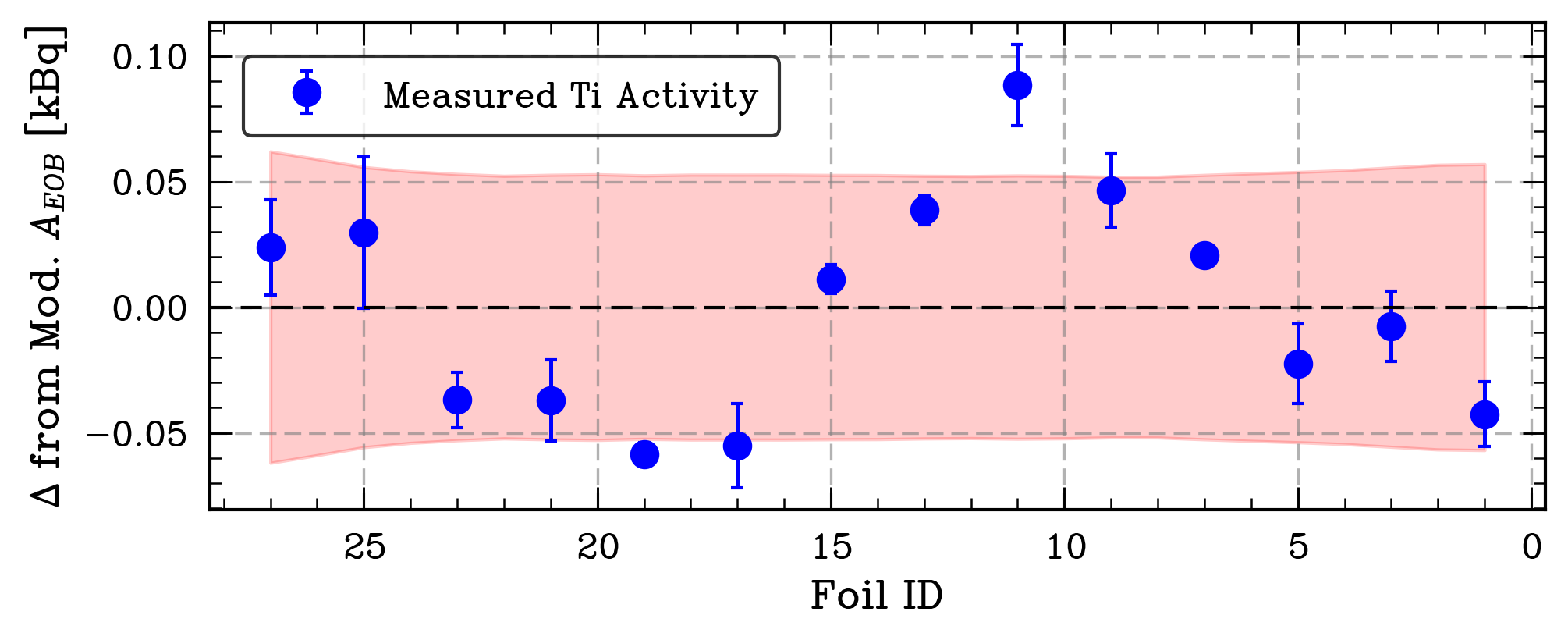}
        \caption{Residual plot for (b).} 
        \label{fig:sub2_res}
    \end{subfigure}
    
    \vspace{0.1cm} 

    \begin{subfigure}{0.47\textwidth}
        \includegraphics[width=\linewidth]{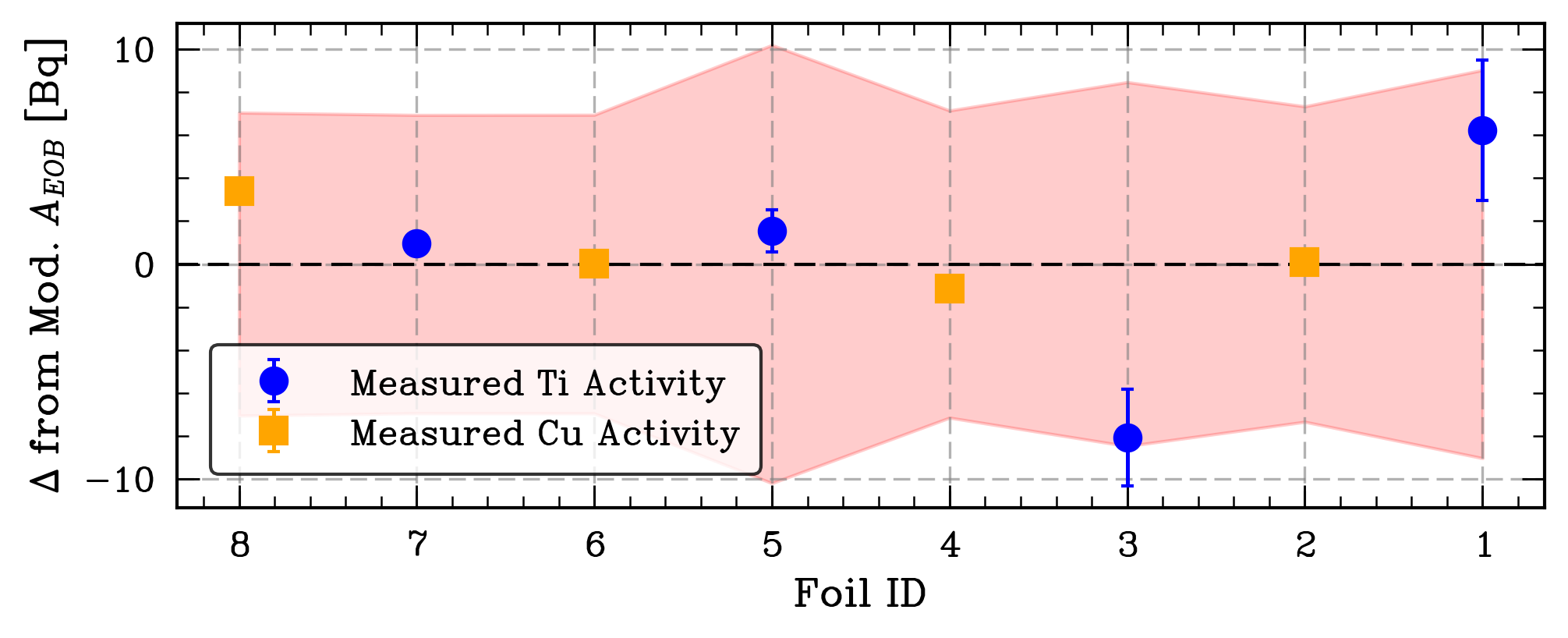}
        \caption{Residual plot for (c).} 
        \label{fig:sub3_res}
    \end{subfigure}
    \hspace{0.35cm}
    \begin{subfigure}{0.47\textwidth}
        \includegraphics[width=\linewidth]{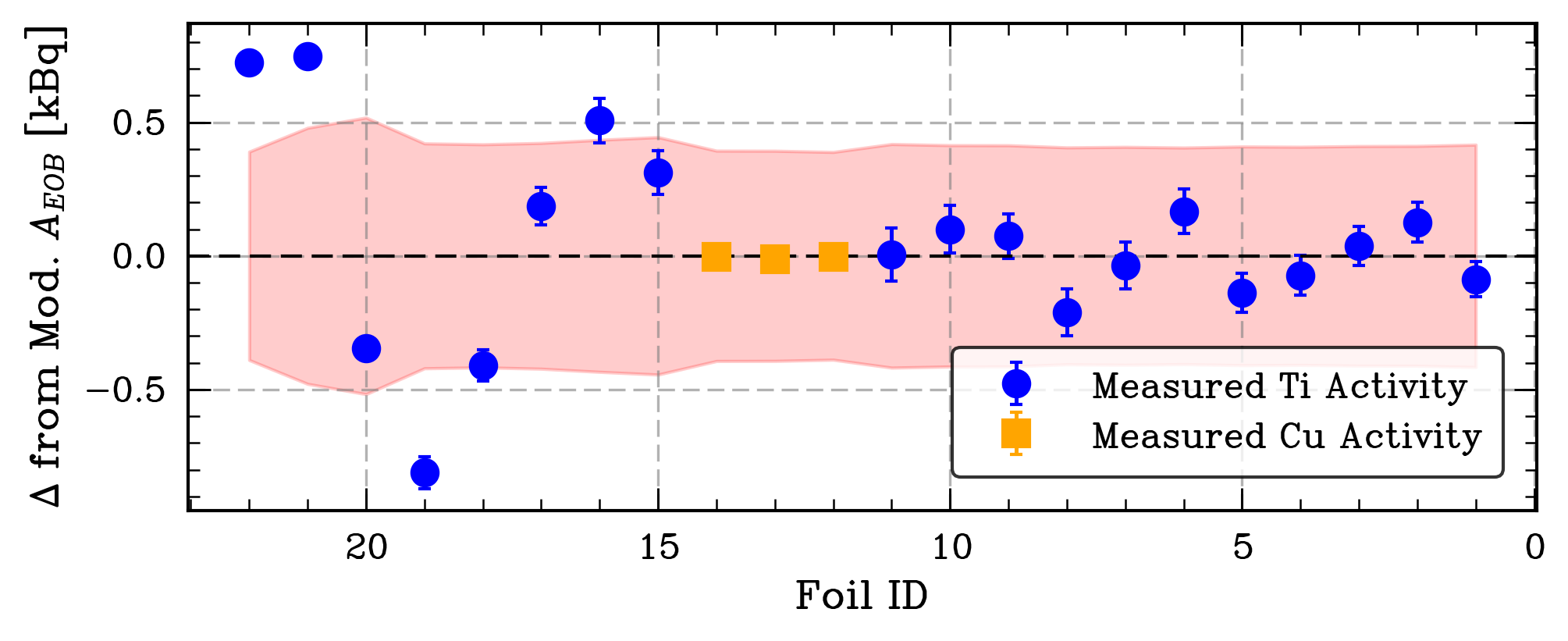}
        \caption{Residual plot for (d).} 
        \label{fig:sub4_res}
    \end{subfigure} 
    \caption{For the four plots in (a) – (d) the residuals of 'measured versus modelled activities' for individual foils are shown in (e) – (h). The y-axis shows the residual in (k)Bq and the x-axis the foil ID (decreasing order reflects depth in the stack). Blue bands indicate model uncertainties. Ti activities are shown as blue markers and Cu as orange markers, where applicable.}
    \label{fig:main_results_residuals}
\end{figure*}

A comparison between the measured activities of each single foil and the corresponding fitted data from the Bayesian model is illustrated in Figures~\ref{fig:sub1}, \ref{fig:sub2}, \ref{fig:sub3}, and \ref{fig:sub4}. In these figures, the red data points represent measured activities of individual foils within the respective stack. The x-axis has been inverted to facilitate direct comparison with Figure~\ref{fig:crosssections_and_stacks}, ensuring that foils positioned towards the right side of the plots correspond to higher interaction energies of the incident protons.
We can see that, in almost all instances, the measured activities fall within the uncertainty range of the fitted data points. Since these fitted points are based on the model’s underlying cross-section, they will also follow the shape of the cross-section function $\sigma(E)$, i.e. the one shown in Figure~\ref{fig:crosssections_and_stacks}. This shape, however, as discussed in the following Section \ref{sec:cs_unc}, is not necessarily representing the true nature of the cross-section. This is why the residuals in Figure~\ref{fig:main_results_residuals} look more systematically scattered than normally distributed around the modelled values.

\subsection{Cross-Section Uncertainty} \label{sec:cs_unc}

\noindent For the results presented above, the IAEA-recommended cross-section (see \cite{hermanne_reference_2018}) for the $^\text{nat}\text{Ti}(\text{p,x})^{48}\text{V}$ monitor reaction was used. However, as highlighted in previous studies (e.g. \cite{cervenak_new_2020}), alternative approaches to fitting the available cross-section data may yield improved consistency or accuracy for our model, as this data plays a central role in our analysis. Following a similar methodology to that from \citep{cervenak_new_2020}, we performed a custom spline fit to selected experimental data from the EXFOR database to construct an alternative cross-section for the \TiV reaction. This custom cross-section is compared to the IAEA-recommended data in Figure~\ref{fig:cross section comparison}. The two cross-sections have their main differences in the energy region from \MeV{9} up to \MeV{15}, as well as in the drop-off region around \MeV{6.5}. 
To assess whether our beam energy measurement method is sensitive to the specific choice of cross-section, we repeated the fitting procedure using both cross-section datasets. A comparison of the resulting best-fit parameters and their uncertainties is provided in Table~\ref{tab:cs comparison}.

\begin{table*}[h]
    \centering
    \begin{tabular}{|>{\centering\arraybackslash}p{3cm}||>{\centering\arraybackslash}p{3cm}|>{\centering\arraybackslash}p{3cm}|}
    \hline
    \rowcolor[gray]{0.9} \textbf{Measurement} & \textbf{IAEA recomm.} & \textbf{Custom Fit} \\
    \hline
    BTL Energy \raisebox{.5pt}{\textcircled{\raisebox{-.9pt} {\small 1}}} & $18.03 \pm 0.09$ \qty{}{\mev} & $18.12 \pm 0.07$ \qty{}{\mev}\\
    \hline
    After Scatterer \raisebox{.5pt}{\textcircled{\raisebox{-.9pt} {\small 2}}} & $15.54 \pm 0.12$ \qty{}{\mev} & $15.51 \pm 1.11$ \qty{}{\mev}\\
    \hline
    Cell Level \raisebox{.5pt}{\textcircled{\raisebox{-.9pt} {\small 3}}} & $\,\,\, 8.14\pm 0.29$ \qty{}{\mev} & $\,\,\, 8.39 \pm 0.20$ \qty{}{\mev}\\
    \hline
    STS Energy \raisebox{.5pt}{\textcircled{\raisebox{-.9pt} {4}}} & $17.14 \pm 0.13$ \qty{}{\mev} & $17.21 \pm 0.06$ \qty{}{\mev}\\
    \hline
    \end{tabular}
    \caption{Comparison of best-fit energy parameters obtained when using the two different \TiV cross-sections from Figure~\ref{fig:cross section comparison} as input to the Bayesian model.}
    \label{tab:cs comparison}
\end{table*}

\noindent We see that for all four cases the results agree within their respective uncertainty, and the change of the best-fit central energy does not change by more than 3\% with the new choice of cross-section data. As we can see in Figure~\ref{fig:cross section comparison}, the two utilized cross-sections differ most in the energy regions of 5.5~–~\qty{7.5}{\mev} and 9 – \qty{12}{\mev} energy regions. As shown in Figure~\ref{fig:crosssections_and_stacks}, the cell level energy measurement is the only measurement of the four that does not contain an irradiated Ti sample outside of the pivotal energy region. Thus, it is clear that this measurement is the most significantly impacted by the different choice of cross-section among the four.
\subsection{Sample Size Reduction} \label{sec:SampleSizeReduction}

\begin{figure}[h]
    \centering
    \includegraphics[width=1.0\linewidth]{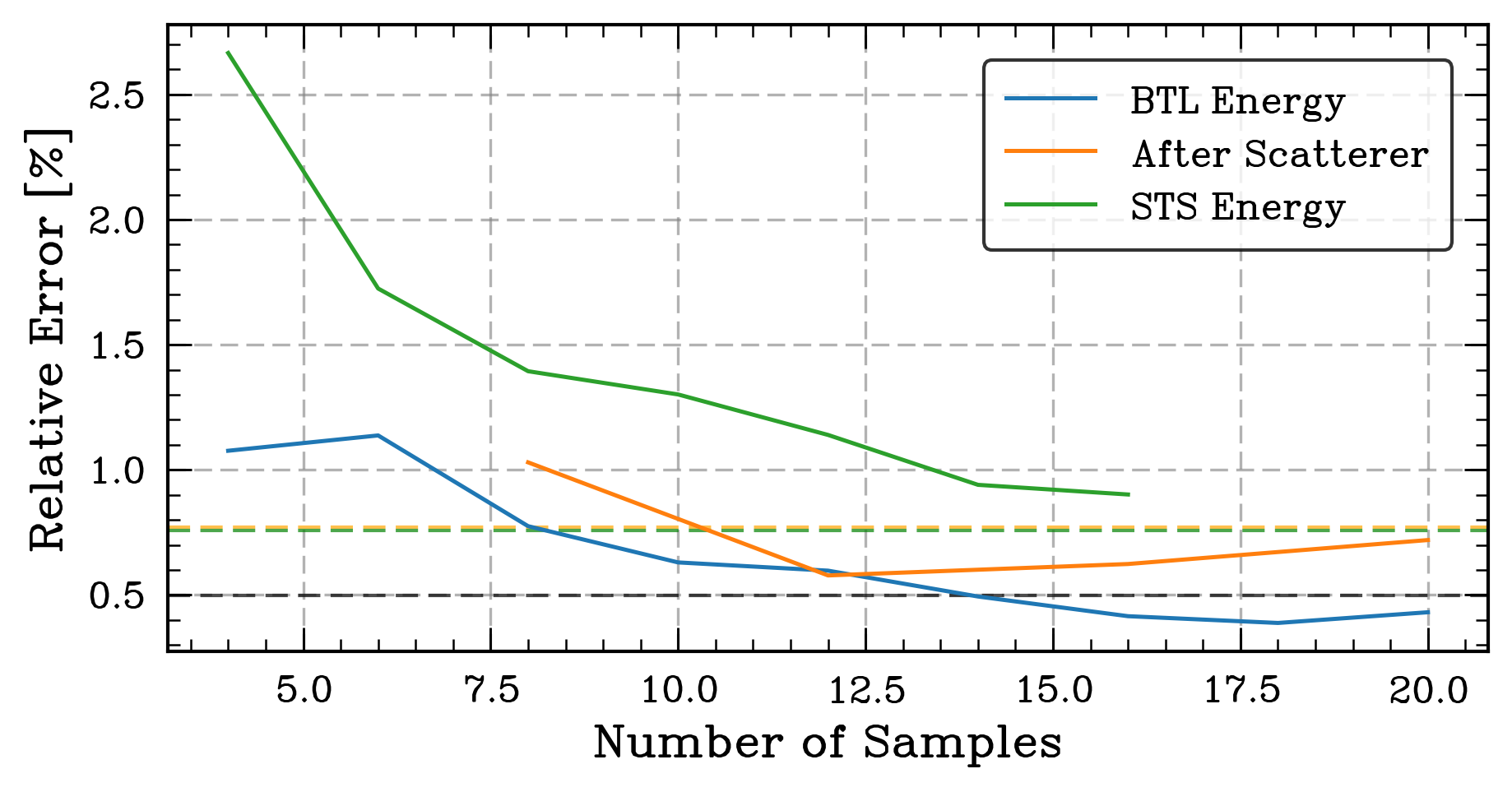}
    \caption{Relative error of the Bayesian fit model for three cases using reduced datasets, where activity data points from pairs of foils are removed simultaneously from the high- and low-energy ends of the stack. Dashed lines represent the relative errors obtained with all available foils. The relative errors of the ‘After Scatterer’ (\raisebox{.5pt}{\textcircled{\raisebox{-.9pt} {\scriptsize 2}}}) and ‘STS Energy’ (\raisebox{.5pt}{\textcircled{\raisebox{-.9pt} {4}}}) measurements are nearly identical, with their respective dashed lines almost entirely overlapping.}
    \label{fig:sample_reduction_rel_err}
\end{figure}

A key question arising from the measurement is how many irradiated samples are required for the method to yield a reliable energy determination with sufficiently low uncertainty, i.e. the same uncertainty as the one achieved with the full sample size. To investigate this, the relative error of the Bayesian fitting procedure has been analysed for reduced datasets of the measurements of \raisebox{.5pt}{\textcircled{\raisebox{-.9pt} {\small 1}}}, \raisebox{.5pt}{\textcircled{\raisebox{-.9pt} {\small 2}}}, and \raisebox{.5pt}{\textcircled{\raisebox{-.9pt} {4}}}. From the complete datasets shown in Figures~\ref{fig:sub1}~-~\ref{fig:sub4}, data points have been removed pairwise from the outer ends, that is, the foils corresponding to the highest and lowest energies were excluded. The resulting relative errors of the modelled beam energies are plotted in Figure~\ref{fig:sample_reduction_rel_err}. Although this investigation is intended as an initial exploration rather than a definitive determination of the optimal sample size, the observed trend indicates that equivalent relative error could be achieved with sample sizes smaller than that of the complete datasets. As 

\section{Conclusion and Outlook}

The stacked foil method has been investigated as a reliable method to measure the proton beam energy. Our studies show that this method can be well applied to proton beams in the energy range between \MeV{8} and \MeV{19}. For beam energies above \MeV{13} it can be enough to irradiate a stack of about ten Ti foils of \qty{25}{\micro \meter} thickness to get a good measurement of the central beam energy of the beam. For energy measurements on the lower end of the mentioned energy range, it becomes necessary to use a second reaction cross-section besides the \TiV one to increase the accuracy of the measurement. Of course, depending on the proton energy, other (monitor) cross-sections can be used as the main measurement attribute. Possible reactions include the following: The $^{27}$Al(p,x)$^{22}$Na reaction for the 30-\MeV{60} energy range can be coupled with $^{27}$Al(p,x)$^{24}$Na for the 30-\MeV{100} range, $^{nat}$Ti(p,x)$^{46}$Sc for measurements around \MeV{50}, $^{nat}$Ni(p,x)$^{57}$Ni for around \MeV{35}, and $^{nat}$Cu(p,x)$^{58}$Co for around \MeV{70}. What makes the method particularly useful is that it does not depend on a current-on-target measurement. This allows for measurements of beam energies in 'non-standard set-ups' such as in the case of the energy measurement with set-up \raisebox{.5pt}{\textcircled{\raisebox{-.9pt} {\small 3}}}. In this configuration, the current-on-target measurement would have been unreliable, as the target station operated in an independent vacuum region whose pressure could not be monitored with sufficient precision. More generally, the simplicity of the method allows its use in any accelerator laboratory, as it requires no special beamline equipment or targets, and only relies on standard gamma-spectroscopy.

For this reason, the stacked foil method may also serve as an independent verification of the current-on-target in configurations where, for example, the use of a Faraday cup is not feasible. In the Bayesian framework, the integrated current on target $Q$ is treated as a free parameter - just like the beam energy $E_0$ - and can therefore be compared to a direct current measurement for consistency.

To minimise the workload needed to characterise the beam energy with the described stacked foil technique, in addition to decreasing the sample size, one could also consider changing the foil thickness. With thicker foils, a wider energy range of the cross-section could be covered while using fewer foils. Even a varying foil thickness is conceivable; for example, the steep region of the cross-section can be covered with thin foils, while the flat region can be covered with a few thicker foils. 
A sufficient number of data points has to be ensured, in particular in the energy regions with steep cross-section gradients. Furthermore, when increasing the thickness of the foils it is important not to violate the basic assumptions of the method, i.e.\ negligible change of the fluence (Equation \eqref{eq:A_layer_2} is valid) and negligible self-absorption for the HPGe activity measurements. 
The optimal combination of foil thickness and number of layers can therefore be tailored to further improve the effectiveness of the method. However, the most suitable choice depends on the specific requirements of the individual measurement (e.g. energy range, available irradiation time, detector sensitivity) and cannot be universally prescribed for the stacked foil method in general.
\section*{CRediT authorship contribution statement}
\textbf{Alexander~Gottstein}: Conceptualization, Methodology, Investigation, Data curation, Formal analysis, Visualization, Writing – original draft. \textbf{Lorenzo~Mercolli}: Methodology, Formal analysis, Software, Visualization, Writing – original draft. \textbf{Eva~Kasanda}: Investigation, Data curation, Validation, Writing – review and editing. \textbf{Isidre~Mateu}: Conceptualization, Methodology, Formal analysis, Investigation, Data curation, Writing – review and editing. \textbf{Lars Eggimann}: Investigation, Data curation. \textbf{Elnaz~Zyaee}: Methodology, Investigation. \textbf{Gaia~Dellepiane}: Methodology, Investigation, Data curation, Writing – review and editing. \textbf{Pierluigi~Casolaro}: Investigation, Methodology, Visualization, Writing – review and editing. \textbf{Paola~Scampoli}: Writing – review and editing. \textbf{Saverio~Braccini}: Conceptualization, Methodology, Resources, Supervision, Project administration, Funding acquisition, Writing – review and editing.

\section*{Declaration of competing interest}
The authors declare that they have no known competing financial
interests or personal relationships that could have appeared to influence
the work reported in this paper.

\section*{Acknowledgment}

We are grateful for the contributions of the LHEP engineering and technical staff (Silas Bosco and Finn Tschan, in particular for this work). This research was partially funded by the Swiss National Science Foundation (SNSF). Grants: 
IZURZ2\_224901 and CRSII5\_180352.

\section*{Data availability}
Data will be made available on request.

\bibliographystyle{elsarticle-num-names}
\biboptions{numbers,sort&compress}
\bibliography{references}

\end{document}